\documentclass[12pt]{article}
\usepackage{amsmath,mathtools,amsthm,amssymb,euscript,epsf,epsfig,color,graphicx}
\usepackage{array}
\usepackage{fancybox}
\usepackage{hyperref} 
\usepackage{scalerel}

\newcommand{\be}{\begin{equation}}
\newcommand{\ee}{\end{equation}}
\newcommand{\bea}{\begin{eqnarray}\displaystyle}
\newcommand{\eea}{\end{eqnarray}}

\makeatletter
\@addtoreset{equation}{section}
\makeatother

\def\one{{\hbox{ 1\kern-.8mm l}}}
\def\zero{{\hbox{ 0\kern-1.5mm 0}}}

\def\mZ{ \mathbb{Z}}

\def\tr{ {\rm{tr}}}

\def\mN{ \mathbb{N} } 

 \def\cH{{\cal H}}

\def\cP{{\cal P}}  
\def\cS{{\cal S}}  
  
 \def\cZ{{\cal Z}}

\def\Sym{ {\rm{ Sym }} } 

\def\Exp{ { \rm{Exp} } }

\begin{document}

\begin{flushright}
QMUL-PH-23-32 \\
DIAS-STP-23-22
\end{flushright}

\bigskip

\begin{center}

 
 {\Large \bf   Gauged permutation invariant matrix quantum mechanics: Partition functions 
\\
 }
 \medskip

\bigskip

{\bf Denjoe O'Connor}$^{a,*}$, {\bf Sanjaye Ramgoolam}$^{b , c ,\dag}  $

\bigskip

$^a${\em School of Theoretical Physics } \\
{\em Dublin Institute of Theoretical Physics,
10 Burlington Road, Dublin 4, Ireland } \\
\medskip
$^{b}${\em School of Physics and Astronomy} , {\em  Centre for Research in String Theory}\\
{\em Queen Mary University of London, London E1 4NS, United Kingdom }\\
\medskip
$^{c}${\em  School of Physics and Mandelstam Institute for Theoretical Physics,} \\   
{\em University of Witwatersrand, Wits, 2050, South Africa} \\
\medskip
E-mails:  $^{*}$denjoe@stp.dias.ie,
\quad $^{\dag}$s.ramgoolam@qmul.ac.uk

\begin{abstract} 
The Hilbert spaces of matrix quantum mechanical systems with $N \times N$  matrix degrees of freedom $ X $ have been analysed recently in terms of $S_N$ symmetric group elements $U$  acting as $X \rightarrow U X U^T $. Solvable models have been constructed  uncovering partition algebras as hidden symmetries of these systems. The solvable models include an 11-dimensional space of matrix  harmonic oscillators, the simplest of which is the standard matrix harmonic oscillator with $U(N)$ symmetry. The permutation symmetry is realised as gauge symmetry in a path integral formulation in a  companion paper. With the simplest matrix oscillator Hamiltonian subject to gauge permutation symmetry, we use the  known result for the micro-canonical partition function to derive the canonical partition function. It is expressed as a sum over partitions of $N$ of products of  factors which depend on elementary number-theoretic properties of the partitions, notably the least common multiples and greatest common divisors  of pairs of parts appearing in the partition. This formula is recovered using the Molien-Weyl formula, which we review for convenience. The Molien-Weyl formula is then used to generalise the formula for the canonical partition function to the 11-parameter permutation invariant  matrix harmonic oscillator. 

\end{abstract}

\end{center}


\newpage

\tableofcontents

\section{ Introduction  } 

Following the long tradition of diverse applications of traditional matrix models with continuous symmetries to real world data ranging across nuclear physics, condensed matter physics and statistical finance (see e.g. \cite{Wigner,RMT-Quantum,EdelWang}), permutation invariant matrix models with Gaussian action have been classified and applied to the statistical analysis of ensembles of matrices representing real-world data \cite{LMT,PIGMM,PIG2MM,HSLNF,PIGMFC}. 
Permutation invariant sectors of quantum mechanical systems with matrix degrees of freedom have been characterised using partition algebras and solvable models have been described \cite{PIMQM} by adapting techniques based on Schur-Weyl duality for matrix systems with continuous symmetries \cite{CJR,KR,BHR,dMK1}. The development of these quantum systems is motivated by the  search for  similarities in the physics between models with permutation symmetry and those models which have known holographic duals such as \cite{bfss,bmn}. This can inform the search for holographic duals of matrix quantum mechanics models with finite group symmetries.   In  \cite{GPIMQM-PI} permutation invariance in matrix quantum mechanics has been realised as a gauge symmetry. The path integral has been described as a limit of a lattice construction and, for the case of the matrix harmonic oscillator, the Molien-Weyl formula for invariants has been obtained from the path integral. 

In this paper the first system we consider is the standard matrix harmonic oscillator  with Hamiltonian 
\bea 
 H = \sum_{ i , j =1}^{ N } A^{ \dagger}_{ ij} A_{ ij} 
\eea 
There is an  action on the matrix harmonic oscillator by permutations $\sigma $ in the symmetric group of all permutations 
$\sigma : \{ 1, 2, \cdots  ,  N \} \rightarrow \{ 1, \cdots , N \}$ : 
\bea \label{permaction} 
A^{ \dagger}_{ ij} \rightarrow A^{ \dagger}_{ \sigma (i) \sigma(j) } 
\eea
Defining  the permutation matrix 
\bea 
( U_{\sigma } )_{ ij}  = \delta_{ i , \sigma^{-1}  (j) } 
\eea
we have 
\bea 
( ( U_{\sigma } )^{T})_{ ij}  = \delta_{ j , \sigma^{-1}  (i) }  = \delta_{ i , \sigma ( j ) } 
\eea
The action \eqref{permaction} can be expressed in terms of matrix multiplication as 
\bea 
A^{ \dagger} \rightarrow U_{\sigma }  A^{ \dagger} U^T_{ \sigma} 
\eea
The oscillators transform as $V_N \otimes V_N$ where $V_N$ is the natural representation of $S_N$.  $U_{ \sigma }$ is real and  $A_{ij}$ transform in the same way. In the subspace of the Fock space with $k$ oscillators, we have polynomials in $A^{ \dagger} $ of degree $k$. The problem of projecting degree $k$ polynomials in $N \times N$ matrices to the  $S_N$ invariant subspace was considered in \cite{LMT} and a formula $\cZ ( N , k )$  
(reproduced here as \eqref{LMTform})  was derived by using projectors in the group algebras of $S_N $ and $S_k$ along with  characters
in $V_N^{ \otimes k } $.  In section \ref{FirstDerBosPart} we use this micro-canonical formula to derive the canonical partition function for the gauge-invariant subspace of the Hilbert space 
\bea 
\cZ ( N ,  \beta ) = \tr (  e^{ - \beta H}  P_0 ) = \sum_{  k  }  e^{ - \beta k } \cZ ( N , k )  \equiv \sum_{  k  }  x^k  \cZ ( N , k ) = \cZ ( N , x )   
\eea 
We have defined $x = e^{ - \beta }$. The projector $P_0$ projects to the $S_N$ invariant subspace. Our first main result is to obtain a formula 
for $\cZ ( N , x  ) $ as a sum over partitions of $N$ \eqref{partitionSum} consisting of an inverse symmetry factor 
$\Sym ~p $ associated with the partition,   multiplied by a factor $ \cZ ( N , p , x )$ \eqref{MainProp}. This latter is a product of factors whose structure depends on elementary number theoretic properties of the parts appearing in the partition.  These parts of the partition are cycle lengths of the permutations belonging to the conjugacy class in $S_N$ associated with the partition. Specifically if the partition $p$ consists of  parts $a_i $ appearing with multiplicities $p_i$, then $ \cZ ( N , p , x )$ is  a product labelled by the parts $a_i$ and another product labelled by pairs $( a_i, a_j)$ of parts in the partition. This second product depends on the least common multiples  $ \hbox{ LCM}  ( a_i , a_j ) \equiv  L ( a_i , a_j )$ of the pairs and the greatest common divisors $\hbox{ GCD} ( a_i , a_j ) \equiv G ( a_i , a_j )$ of the pairs. 

There is an interesting formula, called the Molien-Weyl formula  (see for example \cite{TextBookMW} for a textbook discussion),  for the generating function of bosonic Fock space states invariant under a symmetry group $G$,  which is expressed as an inverse determinant involving the matrix representing the action of $G$ on the bosonic oscillators. In section \ref{sec:RevMW}, 
we recall this formula and, for completeness, give a derivation in section  \ref{MWderBos}. The derivation is generalised in section \ref{MWderWghtBos}   to the case where the bosonic oscillators transform according to a direct sum of representations, and the Fock space state counting is done with different parameters weighting the different representations. This amounts to having different frequencies for the harmonic oscillators transforming in   the different irreducible  representations. We also generalise the derivation of  section \ref{MWderBos} to the case of Fermionic Fock spaces in section \ref{MWderFerm}.

In section 4, we recover the formula \eqref{MainProp} using the Molien-Weyl formula discussed in  section \ref{sec:RevMW}. 
In section 5,  we use the refined Molien-Weyl formula from section \ref{MWderWghtBos}   to calculate the partition function for the $S_N$  Hamiltonian of the general 11-parameter permutation invariant matrix harmonic oscillator system solved in \cite{PIMQM}. 
The partition function depends on 7 of the 11 parameters and has a structure of a sum over partitions of $N$ weighted by an inverse symmetry factor of the partition along with elementary number theoretic properties of the parts appearing in the partition.  This is the second main result of this paper.

\section{ Bosonic GPIMQM : Partition functions   }\label{FirstDerBosPart}

We find an interesting formula for $S_N$ invariant partition functions with a number theoretic characteristics. It is a sum over partitions of $N$, which correspond to cycle structures of $S_N$ permutations, where the  \hbox{ LCM}  and \hbox{ GCD}  of pairs of cycle lengths play a role.

For permutations $\sigma \in S_N$,  let $U_{ \sigma } $ be the linear operator acting in the natural representation $V_N$ of $S_N$.
Matrix bosonic oscillators $A^{\dagger}_{ i  j } $ with $i , j \in \{ 1, 2, \cdots , N \}$ admit an action of $U_{ \sigma } $ : 
\bea 
A^{ \dagger}_{ ij}  \rightarrow (U_{ \sigma })_{ ik}  A^{ \dagger}_{ kl }  ( U_{ \sigma }^{ T })_{ lj }  
\eea
or in matrix notation 
\bea 
A^{ \dagger}  \rightarrow U A^{ \dagger} U^T \, . 
\eea
The action can also be written as : 
\bea\label{sigbos}  
A^{ \dagger}_{ ij} \rightarrow A^{ \dagger}_{ \sigma (i) \sigma (j) } 
\eea

The dimension of the subspace of the Fock space of these oscillators, at degree $k$, which is invariant under the $S_N$ action has been computed in eqn (B.9) of \cite{LMT}. The discussion in the paper \cite{LMT} is in the context of  polynomial functions of a classical matrix variable $M_{ ij}$  invariant under the action 
\bea 
M_{ ij} \rightarrow M_{ \sigma(i) \sigma (j)  } 
\eea
and the mathematics of this invariant theory question evidently  applies equally well to the same action on bosonic oscillators. 
The dimension of the space of $S_N$ invariants at degree $k$ is  given as a sum of partitions of $N$ and $k$ 
\bea\label{LMTform}  
\cZ ( N , k ) &=&  \sum_{ p \vdash N } \sum_{ q \vdash k } 
{ 1 \over \Sym ~ p } { 1 \over \Sym ~ q } \prod_{i=1}^{k}  \left ( \sum_{ l|i } l p_l  \right )^{ 2 q_i } \cr 
 &=& \sum_{ p \vdash N } { 1\over \Sym ~  p  }  \cZ ( N , p , k ) 
\eea
where 
\bea\label{LMTform1}  
\cZ ( N , p , k )  = \sum_{ q \vdash k }  { 1 \over \Sym ~ q } \prod_{i=1}^{k}  \left ( \sum_{ l|i } l p_l \right  )^{ 2 q_i }
\eea
The partition $p$ of $N$ is described as a list of cycle lengths $l$ of permutations in $S_N$  (parts of the partition) with multiplicities $p_l$ where  $l \in \{ 1, \cdots, N \} $ and $p_l $ are non-negative integers obeying $ \sum_{ l } l p_l = N$. The partition $q$ of $k$ is described as a list of cycle lengths $i$ for permutations in $S_k$ with $i \in \{ 1, \cdots , k \}$ and $q_i$ are non-negative integers obeying $\sum_i i q_i = k $. The symmetry factor $\Sym ~q $ is given by  
\bea 
\Sym ~ q = \prod_{ i =1}^k i^{ q_i} q_i! 
\eea
It is the number of permutations  $\gamma \in S_k$ which obey $\gamma \sigma \gamma^{-1} = \sigma $ for any $\sigma $ having cycle structure $q$.   A useful fact we often use to convert sums over permutations into sums over conjugacy classes is that the number of permutations with cycle structure $q$ is 
\bea 
{ k! \over \Sym ~ q } 
\eea
 
We define the generating function 
\bea 
\cZ ( N , x )  && = \sum_{ k =0 }^{ \infty }  x^k \cZ  ( N , k ) \cr 
&& = \sum_{ k =0 }^{ \infty } \sum_{ p \vdash N } { x^{ k }  \over \Sym ~ p } \cZ  ( N , p ,  k )\cr 
&& = \sum_{ p \vdash N } { 1 \over \Sym ~ p } \sum_{ k }   x^{ k } \cZ  ( N , p ,  k )
\eea
It is also useful to define a generating function for fixed $N $ and fixed partition $p$ of $N$  by summing over $k$
\bea 
\cZ ( N , p , x )   =  \sum_{ k }   x^{ k } \cZ  ( N , p ,  k )
\eea
 $ \cZ ( N , x ) $ can therefore be written as a sum 
\bea\label{partitionSum} 
\cZ ( N , x ) = \sum_{ p \vdash N  } { 1 \over \Sym ~ p } \cZ ( N , p , x )  
\eea

Partitions of $N$ can be described in   the form $ p = [ a_1^{  p_1} , a_2^{ p_2} , \cdots , a_{ K}^{  p_K} ] $, where 
$a_j$ are distinct non-zero parts  with $1\le a_j \le N $ and $p_j $ are positive integers, obeying 
\bea 
N = \sum_{ j =1}^K a_j  p_j 
\eea
It is often useful to think of  the parts as being  ordered as $ a_1 < a_2 < \cdots < a_K $. 
For example for partitions of $4$, we have in this notation 
\bea 
 4 = 4 &\longrightarrow &  p = [ 4 ] \cr 
 4 = 3 + 1 & \longrightarrow &  p = [ 3,1 ] \cr 
 4 = 2 +2 & \longrightarrow  & p = [2^2 ] \cr 
 4 = 2 +1 + 1 & \longrightarrow &  p = [ 2, 1^2 ] \cr 
  4 = 1+1 +1 +1 & \longrightarrow  & p = [ 1^4 ] 
\eea

\noindent 
We will derive the following formula for $\cZ ( N , p ,  x) $ which we refer to as the LCM formula: 
\vskip.2cm 
\noindent 
{\bf Proposition 1: The LCM formula }   \\
\begin{equation}\label{MainProp}  
\boxed{ ~~~~~~
\cZ  ( N , p ,  x) = \prod_{ i =1  }^K  { 1 \over ( 1 - x^{ a_i } )^{ a_i  p_i^2 }   }
 \prod_{ i < j \in \{ 1, \cdots , K \} } { 1 \over ( 1 - x^{ L ( a_i ,   a_j )  } )^{ 2 G ( a_i ,  a_j )    p_i  p_j }   }
 ~~~~~~~ } 
\end{equation} 
$ L ( a_i , a_j ) $ is the least common multiple of $ a_i $ and $ a_j $ and $G ( a_i , a_j ) $ is the greatest common denominator of $a_i , a_j$. 

A   fact relating the  LCM and GCD of a pair of numbers which we will find useful   is that 
\bea 
 a_i  a_j  = L ( a_i, a_j ) G ( a_i , a_j ) 
\eea
This is proved in the equation  \eqref{PGLP}. The expression $ 2 G ( a_i , a_j ) p_i p_j $ in \eqref{MainProp} can also be written as : 
\bea 
2 G ( a_i , a_j ) p_i p_j = {2 a_i a_j p_i p_j \over L ( a_i , a_j ) } 
\eea

We can also present $p$ as $ [ i^{ p_i } ] $ where $i $ are all integers in the set $\{ 1, \cdots , N \}$ and $p_i$ are non-negative integers (possibly zero). In this case we can write 
\bea\label{LCMformula}  
\cZ ( N , p , x )  = \prod_{ i =1  }^N  { 1 \over ( 1 - x^{ i } )^{ i  p_i^2 }   }
 \prod_{ i < j \in \{ 1 , \cdots , N \}  } { 1 \over ( 1 - x^{ L ( i ,   j )  } )^{ 2 G ( i,  j )   p_i  p_j  }   }
\eea
The terms with $p_i=0$ all give factors of $1$ so this immediately reduces to the previous formula \eqref{MainProp}.

Examples of \eqref{MainProp} are as follows. For $N =2$, we have two partitions $p = [2] $ and $p = [1^2]$ : 
\bea 
&& \cZ (  2, [2 ] , x ) = { 1 \over ( 1 - x^2 )^2 } \cr 
&& \cZ ( 2 , [1^2] ,x ) =  {1 \over ( 1 - x )^4 }
\eea
Using \eqref{partitionSum}
\bea
\cZ ( N =2 , x ) = { 1 \over 2 ( 1 - x^2 )^2 } + { 1 \over  2 ( 1 - x )^4 }
\eea
For $ N =3$, there are three partitions $ p = [ 3 ] ,  p  = [2,1] , p =  [1,1,1] =[1^3]  $  and using \eqref{MainProp} we have 
\bea 
&& \cZ ( 3  , [ 3 ] , x ) = { 1 \over ( 1 - x^3 )^3 } \cr 
&& \cZ ( 3  , [ 2,1 ] , x ) = { 1 \over ( 1 - x^2 )^4 ( 1- x )  } \cr 
&&   \cZ ( 3  , [ 1^3  ] , x ) =  { 1 \over ( 1 - x)^9 }
\eea
Using \eqref{partitionSum} then gives 
\bea 
\cZ ( N = 3  , x ) = { 1 \over 3 ( 1 - x^3 )^3 } + { 1 \over  2 ( 1 - x ) ( 1 - x^2 )^4 } + { 1 \over 6 ( 1 -x)^9 } 
 \eea

\subsection{ Proof of LCM formula } 

We will start with  equation \eqref{LMTform1}. The partition $p$ of $N$ 
 is described as $p = [l^{ p_l } ] $ with $\sum_{ l} lp_l = N $. The partition  $q$ of $k $ is described as $q = [i^{ q_i }] $  where $ \sum_{ i } i q_i = k $. 
\bea\label{fststeps}  
 \cZ( N , p ; x ) &  =  &  \sum_{ k =0}^{ \infty }  x^k\sum_{ q \vdash k } { 1 \over \Sym ~ q }\prod_{i=1}^{k}  (  \sum_{ l|i } l p_l  )^{ 2    q_i } \cr 
 & = & \sum_{ k =0}^{ \infty }  x^k \sum_{\substack {  q_1 , q_2 \cdots \\ \sum_{ i } i q_i = k }} { 1 \over { \prod_i i^{q_i} q_i! } } 
\prod_{i=1}^{k}  (  \sum_{ l|i } l p_l  )^{ 2    q_i } \cr 
& =&  \sum_{ q_1 , q_2 , \cdots } { x^{ \sum_i i q_i } \over \prod_{ i } i^{ q_i} q_i! } \prod_{i=1}^{ \infty }  (\sum_{ l|i } l p_l )^{2 q_i }  \cr 
&= &  \prod_{ i } \sum_{ q_i } { 1 \over q_i! }   ( { x^{ i  }\over i }  )^{ q_i}  ( \sum_{ l|i } l p_l  )^{2q_i }   \cr 
& = &   \prod_i  \Exp [ { x^i \over i }  ( \sum_{ l|i } l p_l  )^{2 } ]  \cr 
& =  & \Exp [ \sum_{ i=1}^{ \infty } { x^i \over i } ( \sum_{ l|i } l p_l  )^{2 }  ]
\eea
A minor re-writing gives 
\bea\label{Expsumformula}  
\cZ( N , p ; x ) =  \Exp [ \sum_{ i \in \mathbb{N}  } { x^i \over i } ( \sum_{ l|i } l p_l  )^{2 }  ] 
\eea
where $\mN$ is the set of natural numbers. 

For simplicity we first consider $ p = [ a_1^{ p_1} , a_2^{ p_2} ] $, i.e. partitions with just two  types of parts : $a_1 $ with multiplicity $p_1$ and $a_2$ with multiplicity $p_2$. Define the sets
\bea 
 S_1  && =  \hbox{ set of positive integer multiples of } a_1  = \{ m a_1 : m \in \mN  \}  \cr
 S_2  && = \hbox{ set of positive integer multiples of } a_2 = \{ m a_2 : m \in \mN \}  \cr 
 S_{ 12} && =  \hbox { set of positive integers divisible by both } a_1 \hbox{ and } a_2 \cr 
&&  = \hbox{ set of positive integer multiples of } L ( a_1 , a_2 ) = \{ m L ( a_1 , a_2 ) : m \in \mN \}  \cr 
&& 
\eea
Note that 
\bea 
S_{12} \subset S_1 \cr 
S_{ 12} \subset S_2  
\eea
Consider the expression 
\bea 
 \sum_{ i=1}^{ \infty } { x^i \over i } ( \sum_{ l|i } l p_l  )^{2 }
\eea
appearing in the exponent of \eqref{Expsumformula}. Given that $l$ is being summed over divisors of $i$ and the only non-zero $l$'s for $ p = [ a_1^{ p_1} , a_2^{p_2} ] $ are $a_1 $ and $a_2$, it follows that the powers $i$  appearing in this sum belong to a subset $\cS$ of $\mN $ which is  
\bea 
\cS  =  S_1 \cup S_2 \subset \mN   
\eea
It is also useful to note that 
\bea 
 S_1 \cap S_2 = S_{ 12} 
\eea 

For all elements $i = m a_1 \in S_1$, the expression  $( \sum_{ l|i } l p_l  )^{2 }$ contains $a_1^2 p_1^2$ plus possibly additional terms. 
For all elements $ i = m a_2 \in S_2$, the same expression contains $a_2^2p_2^2$ plus possibly additional terms. 
For the elements in $ i = m L ( a_1 , a_2 ) \in S_{ 12}$, the expression evaluates to $ ( a_1 p_1 + a_2 p_2)^2 =  ( a_1^2 p_1^2 + a_2^2 p_2^2 + 2 a_1 a_2 p_1 p_2 )$. 
We can therefore write  the sum over $\mathbb { N } $ in \eqref{Expsumformula}  as a sum over $S_1 $ with weight $a_1^2 p_1^2$ and a sum over  $S_2$ with weight $a_2^2 p_2^2$, along with a sum over $S_{ 12}$ with weight $ 2 a_1 a_2 p_1 p_2 $ (since $S_1 \cap S_2 = S_{12} $, the contributions $a_1^2p_1^2 + a_2^2p_2^2$ to the coefficient of $x^{ m L ( a_1 , a_2 )} $ from the expression $( \sum_{ l|i } l p_l  )^{2 }$  are already included in the sums over $S_1 , S_2$).  We can therefore write 
\bea 
&& \sum_{ i \in \mN } { x^i \over i } ( \sum_{ l|i } l p_l  )^{2 }  = \sum_{  m \in \mN  } {  x^{ m a_1 }\over ma_1}   ( a_1 p_1 )^2 
+ \sum_{  m \in \mN  } {  x^{ m a_2 }\over ma_2}   ( a_2 p_2 )^2 + \sum_{ m \in \mN }  {  x^{ m L ( a_1, a_2 )   }\over m L ( a_1,  a_2)  }
 ( 2 a_1 a_2 p_1 p_2 ) \cr 
&& =    \sum_{  m \in \mN  } {  x^{ m a_1 }\over ma_1}   ( a_1 p_1)^2 
+ \sum_{  m \in \mN  } {  x^{ m a_2 }\over ma_2}   ( a_2 p_2 )^2 + \sum_{ m \in \mN }  {  x^{ m L ( a_1, a_2 )   }\over m   }
 ( 2 G ( a_1,  a_2)  p_1 p_2 ) \cr
 &&   
\eea
In the last line we used 
\bea\label{GLP}  
 { a_1 a_2 \over L ( a_1 , a_2 ) } =  G ( a_1 , a_2 ) 
\eea
To see this let us denote $ h = G ( a_1 , a_2 ) $ and $ a_1 = h \hat a_1, a_2 = h \hat a_2$ where $ G ( \hat a_1 , \hat a_2 ) =1 $. 
Then 
\bea\label{PGLP}  
 L ( a_1 , a_2 ) = L ( h \hat a_1 , h \hat a_2 ) = h \hat a_1 \hat a_2  = { a_1 a_2 \over h } = { a_1 a_2 \over G ( a_1 , a_2 ) } 
 \eea 
 which proves \eqref{GLP}.

Using these facts we can write 
\bea\label{twoparts}  
&& \cZ( N , p ; x ) \cr 
      && = \Exp [ \sum_{ m =1  }^{ \infty } { x^{ m a_1  }  \over  m a_1   } (  a_1   p_1  )^{2 }  ] 
      \Exp [ \sum_{ m =1  }^{ \infty } { x^{ m a_2  }  \over m  a_2  } (  a_2   p_2  )^{2 }  ]   \Exp [ \sum_{ m =1  }^{ \infty } { x^{ m  L (  a_1 , a_2 )    }  \over m L ( a_1,  a_2)    } 
      ( 2  a_1 a_2     p_1  p_2  )   ] \cr
      && = \Exp [ \sum_{ m =1  }^{ \infty } { x^{ m a_1  }  \over   m   }  ( a_1   p_1^2 )   ] 
      \Exp [ \sum_{ m =1  }^{ \infty } { x^{ m a_2  }  \over  m   } (  a_2   p_2^{2 })   ]   \Exp [ \sum_{ m =1  }^{ \infty } { x^{ m L ( a_1 , a_2 )    }  \over    m } 
      ( 2   G ( a_1 , a_2 ) p_1  p_2  )   ]  \cr 
      && = \Exp [ - a_1 p_1^2 \log ( 1 - x^{ a_1}  ) - a_2  p_2^2 \log ( 1 - x^{ a_2}  )  - 2  p_1  p_2 G ( a_1 , a_2)  \log ( 1 - x^{ L ( a_1, a_2) } ) \cr 
      &&  = { 1 \over ( 1 - x^{ a_1} )^{ a_1  p_1^2 }} { 1 \over ( 1 - x^{ a_2} )^{ a_2  p_2^2 } }
          { 1 \over ( 1 - x^{ L ( a_1 ,  a_2)  } )^{ 2 G ( a_1 , a_2 )   p_1  p_2 } }
\eea

\noindent 
{ \bf Remarks } 
\begin{enumerate} 
\item We  observe that the above result and derivation also  apply  when one of $a_1 $ or $a_2$ is a multiple of the other. Without loss of generality, take $a_2$ to be $na_1$ for some positive integer $n$.  Then $ S_2 \subset S_1$, $ L ( a_1, a_2 ) = a_2 $,  $G ( a_1 , a_2 ) = a_1 $ and $S_{ 12} = S_2$. The formula in \eqref{twoparts} then simplifies to 
\bea 
\cZ( N , p ; x )  && = { 1 \over ( 1 - x^{ a_1} )^{ a_1  p_1^2 }} { 1 \over ( 1 - x^{ a_2} )^{ a_2  p_2^2 } }
          { 1 \over ( 1 - x^{ a_2 } )^{ 2 a_1   p_1  p_2 } }  \cr 
          && = { 1 \over ( 1 - x^{ a_1} )^{ a_1  p_1^2 }} { 1 \over ( 1 - x^{ a_2} )^{ a_2  p_2^2 + 2 a_1 p_1 p_2  } } \cr 
&& ~~
\eea

\item Another simplification is the case where $ G ( a_1 , a_2 )  =1 $. In that case, $ L ( a_1 , a_2 ) = a_1 a_2 $ and the formula \eqref{twoparts} simplifies to 
\bea 
\cZ( N , p ; x )  = { 1 \over ( 1 - x^{ a_1} )^{ a_1  p_1^2 }} { 1 \over ( 1 - x^{ a_2} )^{ a_2  p_2^2 } }
          { 1 \over ( 1 - x^{ a_1 a_2 } )^{ 2   p_1  p_2 } }
\eea
\end{enumerate} 

The derivation above  extends to the case of general partitions of the form \\
 $ p =  [ a_1^{ p_1 }  , a_2^{ p_2} , \cdots , a_K^{ p_K } ] $. In this case  the sum $ ( \sum_{ l|i } l p_l  )^{2 } $ appearing in \eqref{Expsumformula} will contain, depending on the choice of $i$, terms of type $a_{ \alpha }^2p_{\alpha}^2 $ for $ \{ \alpha  \in \{ 1, 2 , \cdots , K \} \}$,  as well as terms of the type $2a_{\alpha }  a_{\beta }  p_{\alpha}  p_{\beta }  $ for pairs
 $\{ \alpha  < \beta  ; \alpha , \beta \in \{ 1, \cdots , K \} \}$. Generalising the observation in the case $p = [ a_1^{ p_1} , a_2^{ p_2} ] $ the only $x^i$ appearing in the sum $ ( \sum_{ l|i } l p_l  )^{2 } $ have powers $i$ belonging to the subset $\cS \in \mN $ where 
 \bea 
 \cS = S_1 \cup S_2 \cdots \cup S_K \subset \mN 
 \eea
where   
 \bea 
 S_{\alpha}    =  \hbox{ set of positive integer multiples of } a_{\alpha}  = \{ m a_{\alpha}  : m \in \mN  \} 
 \eea 
 The term $a_{\alpha}^2 p_{\alpha}^2 $ will appear as a coefficient for $x^i$ for  $ i \in  S_{ \alpha}  \subset \mN $. 
 The cross-terms $2a_{\alpha } a_{\beta }  p_{\alpha }  p_{\beta }  $  will appear in the subset $S_{\alpha \beta } \subset \mN $ 
 \bea 
  S_{ \alpha \beta } && =  \hbox { set of positive integers divisible by both } a_{\alpha } \hbox{ and } a_{\beta }  \cr 
&&  = \hbox{ set of positive integer multiples of } L ( a_{\alpha  } , a_{\beta }  ) = \{ m L ( a_{\alpha }  , a_{\beta}   ) : m \in \mN \}  \cr 
&& 
 \eea
 Taking advantage of the fact that $S_{ \alpha \beta } \subset S_{ \alpha } $ and $ S_{ \alpha \beta }  \subset  S_{ \beta } $ we therefore conclude that 
 \bea 
&& \sum_{ i \in \mN } { x^i \over i } ( \sum_{ l|i } l p_l  )^{2 }  = \sum_{  m \in \mN  } \sum_{ \alpha  =1}^K   { x^{ m a_{\alpha  }} \over ma_{\alpha  } } 
( a_{\alpha}  p_{\alpha} )^2  
+ \sum_{ m \in \mN } \sum_{ \alpha < \beta \in \{ 1, \cdots , K \} } 
  {  x^{ m L ( a_{\alpha } , a_{ \beta }  )   }\over m L ( a_{\alpha  },  a_{ \beta } )  }
 ( 2 a_{\alpha}  a_{\beta } p_{\alpha }  p_{\beta }  ) \cr 
 && ~~
 \eea
 This allows us to obtain 
 \bea 
&&  \cZ ( N , p , x ) = \Exp [ \sum_{ i \in \mN } { x^i \over i } ( \sum_{ l|i } l p_l  )^{2 }  ] \cr 
&& = \Exp [   \sum_{ \alpha  =1}^K   \sum_{  m \in \mN  } { x^{ m a_{\alpha  }} \over m } 
( a_{\alpha}  p_{\alpha}^2 ) 
+ \sum_{ \alpha < \beta \in \{ 1, \cdots , K \} } \sum_{ m \in \mN } 
  {  x^{ m L ( a_{\alpha } , a_{ \beta }  )   }\over m  }
 ( { 2 a_{\alpha}  a_{\beta } p_{\alpha }  p_{\beta }  \over L ( a_{\alpha  },  a_{ \beta })  }   )  ] \cr 
 && = \Exp [   \sum_{ \alpha  =1}^K  \sum_{  m \in \mN  }  { x^{ m a_{\alpha  }} \over m } 
( a_{\alpha}  p_{\alpha}^2 ) 
+ \sum_{ \alpha < \beta \in \{ 1, \cdots , K \} } \sum_{  m \in \mN  } 
  {  x^{ m L ( a_{\alpha } , a_{ \beta }  )   }\over m   }
 ( 2 G ( a_{\alpha} ,  a_{\beta } )  p_{\alpha }  p_{\beta }  )  ] \cr 
 && = \prod_{ \alpha =1}^K \Exp [ - ( a_{\alpha}  p_{\alpha}^2 )  \log ( 1 - x^{ \alpha } ) ] 
       \prod_{ \alpha < \beta  \in \{ 1, \cdots , K \} } \Exp [  - ( 2 G ( a_{\alpha} ,  a_{\beta } )  p_{\alpha }  p_{\beta }  )  \log ( 1 - x^{ L ( a_{ \alpha} , a_{ \beta } ) }  ) ]\cr 
       && =  \prod_{ \alpha =1}^K { 1 \over ( 1 - x^{ a_{\alpha} } )^{  a_{\alpha}  p_{\alpha}^2  }  } \prod_{\alpha < \beta \in \{ 1, \cdots , K \}  }  { 1 \over ( 1 - x^{ L ( a_{ \alpha} , a_{ \beta }} )^{ 2 G ( a_{\alpha} ,  a_{\beta } )  p_{\alpha }  p_{\beta } }  }
 \eea
This completes the proof of the main proposition \eqref{MainProp}. 

\subsection{ Effective-Graphs for graph counting }

It is known that  the counting of permutation invariants at degree $k$ for matrices of size $N$ is equivalent to the counting of directed graphs with $k$ edges and $N $ vertices (see section 2.2 of \cite{PIG2MM}) and in the range $ N \ge 2k$ it is simply the counting of all directed graphs with $k$ edges \cite{PIGMM}. We have found a generating function for the graph counting at arbitrary $k$ for fixed $N$ \eqref{MainProp}. For partitions of $N$  of the form $ [ a_1^{ p_1 } , a_2^{ p_2} \cdots a_K^{ p_K}  ] $, with $p_i$ parts of length $a_i$ which corresponds to cycle lengths of permutations in $S_N$, the  formula contains a product  with factors associated with the cycle lengths $a_i$ and a product with factors associated with pairs  $(a_i,a_j)$. This formula has an interpretation in terms of a weighted  counting of graphs, which may be considered as ``effective graphs'' which produce the generating function for numbers of directed graphs. For a partition $p = [ a_1^{ p_1} , a_2^{ p_2} , \cdots , a_K^{ p_K}   ] $, we consider a labelled graph with $p_1 + p_2 + 
\cdots + p_K  $ nodes. The first $p_1$ nodes are assigned a  weight $a_1$ each, the next $p_2$ are assigned a  weight $a_2$ and so forth. Consider a set of graphs, each consisting of a single  arc joining either a node to itself or a node to another.  Consider the product, one for each arc, of the form  
\bea\label{weights} 
\prod_{ \hbox {\tiny  arcs} } ( 1- x^{L ( a_{ \hbox{\tiny start} } , a_{ \hbox{\tiny  end} } ) }  )^{ -G ( a_{ \hbox{\tiny start}}  , a_{ \hbox{\tiny  end} } ) } 
\eea
and  define a partition function of weighted graphs as a product over  the weighted graphs (defined by the arcs) : 
\bea 
\cZ_{\hbox{ \tiny weighted graphs}  }  ( N , p , x )  = \prod_{ \hbox{ \tiny arcs} } ( 1- x^{L ( a_{ \hbox{\tiny start} } , a_{ \hbox{\tiny  end} } ) }  )^{ -G ( a_{ \hbox{\tiny start}}  , a_{ \hbox{\tiny  end} } ) } 
\eea
The loops joining nodes labelled $a_i$ to nodes labelled  $a_i$ (including self-loops) produce the product 
\bea\label{weightsself}  
\prod_{ i =1 }^{ K }  ( 1- x^{a_i} )^{ -a_i  p_{i}^2} 
\eea 
since $L( a_i , a_i ) = a_i $ there are $p_{a_i}^2$ such arcs. The arcs joining nodes of different weights produce 
\bea\label{weightscross}  
\prod_{ i < j } ( 1- x^{ L(a_i,a_j ) } )^{ - 2 p_i p_j G( a_i , a_j  ) }  
\eea
 since there are $2 p_{i}  p_j $ such arcs. Taking the product over all the weighted graphs we have 
 \bea 
 \prod_{ i =1 }^{ K }  ( 1- x^{a_i} )^{ -a_i p_{a_i}^2} \prod_{ i < j } ( 1- x^{ L(a_i,a_j ) } )^{ - 2 p_i p_j  G( a_i , a_j  ) }  
 \eea
which is the formula \eqref{MainProp}. 
This proves 
\bea 
\cZ_{\hbox{ \tiny weighted graphs}  }  ( N , p , x )  = \cZ ( N , p , x ) 
\eea 

Figure 1 illustrates the  effective graphs used to produce the generating function. 
The summand $Z ( N , p , x ) $ of $ Z ( N , x ) $ for the case $N=3, p = [2,1]$ is constructed from the effective graph with a node labelled $2$ and a node labelled $1$. The different possible 1-edge directed graphs  are shown in Figure 1. For the first two, we apply the formula  \eqref{weightsself}, while for the last two we apply \eqref{weightscross} to arrive at the function on the right. Taking the product over the four graphs leads to 
\bea 
Z ( N =3, p = [2,1] , x ) = { 1 \over ( 1 - x^2)^4 ( 1- x ) } 
\eea

\begin{figure} 
\includegraphics[scale=0.4]{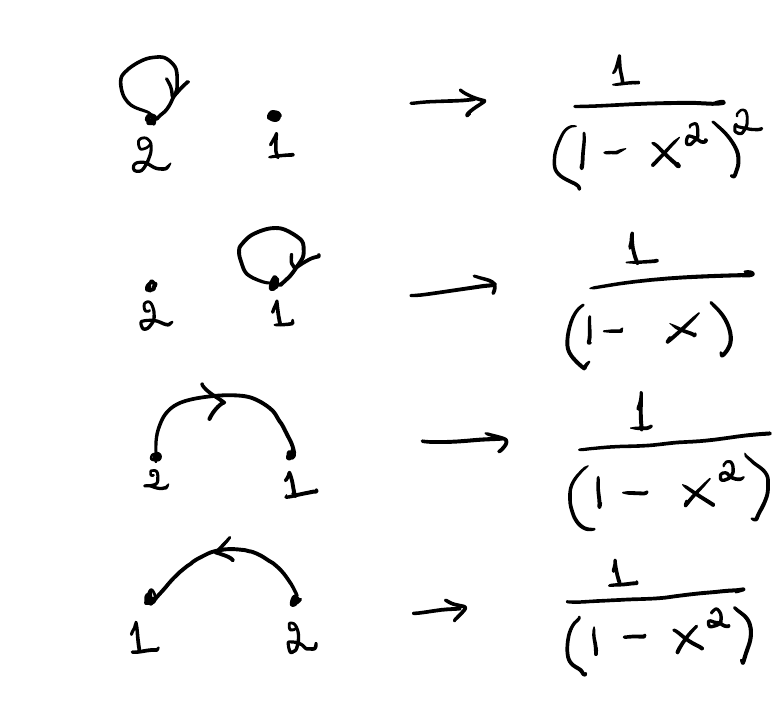}\label{fig1}  
\caption{Diagrams of effective graphs for the calculation of $ Z ( N = 3 , p = [2,1] , x )$}
\end{figure} 

 As a corollary of this discussion, note that 
\bea 
\cZ ( N , p , x ) = \prod_{ i , j \in \{ 1 , \cdots , K \}  } { 1 \over ( 1 - x^{ L ( a_i , a_j ) } )^{ G ( a_i , a_j )}}
\eea
The case where $i = j$ leads to the first product in \eqref{MainProp} while the terms $ i \ne j $ lead to the second product in \eqref{MainProp}.

\section{ Review : Derivation of Molien-Weyl Determinant formula }\label{sec:RevMW}

A standard result which is very useful in the computation of dimensions of subspaces of Fock spaces, or of spaces of polynomials,  invariant under a group action is the Molien-Weyl formula. A textbook presentation and derivation is in \cite{TextBookMW}. In a physics context, this formula has been used for example in \cite{PP1,PP2}. In this section, in the interest of a self-contained presentation, we will give a proof of the Molien-Weyl determinant formula for the generating function of dimensions of state spaces for bosonic Fock spaces using permutations and projector techniques which are used elsewhere in the paper. An alternative derivation proceeds by diagonalising the matrix for $g$  \cite{TextBookMW}. The derivation of the Molien-Weyl determinant formula from the gauged bosonic matrix quantum mechanics path integral is given in \cite{GPIMQM-PI}. The permutation and projector techniques admit an immediate generalization to the case of  Fock spaces defined using oscillators transforming according to a direct sum of irreducible representations, with the different representations weighted by different parameters. This is described in section \ref{MWderWghtBos}. The generalisation to fermionic Fock spaces is  described in Section \ref{MWderFerm}.

\subsection{ Molien-Weyl determinant formula derivation  : Bosonic case}\label{MWderBos} 

Consider a finite group $G$ and a set of bosonic oscillators $A^{ \dagger }_{ m } $  transforming in the 
 representation $V$.  The state space of $k$  bosonic oscillators is the symmetrized $k$-fold tensor power 
 $\Sym^k ( V ) $. The generating function of invariants in the $\Sym^k ( V ) $ is defined as 
\bea\label{bospartfun}  
\cZ_{ {\rm sym} }  ( G , V ; x  )  = \sum_{ k =0}^{ \infty } x^k  {\rm tr}_{ \Sym^k ( V ) } ( P_0 )
\eea 
where the projector to $G$-invariants is 
\bea\label{projGinvts}  
P_0 = {1\over |G|} \sum_{ g \in G } g 
\eea
We therefore have
\bea\label{bospartfun1} 
 \cZ_{ {\rm sym} }  ( G , V ; x  ) &=&  { 1 \over |G| }  \sum_{ g \in G } \sum_{ k =0}^{ \infty } x^k   { \rm tr}_{ \Sym^k ( V ) } ( g ) \cr 
 & = & { 1 \over |G|  } \sum_{ g \in G }  \sum_{ k =0}  { x^k \over k! }   \sum_{ \tau \in S_k } \tr_{ V^{ \otimes k } } ( g^{ \otimes k } \tau ) 
\eea
To calculate the trace in the last line, introduce a basis $e_{ i } $ for $V$, with $   1 \le i \le {\rm Dim}  V $. The permutation $\tau $ viewed as a map $\tau : \{ 1, 2, \cdots , k \} \rightarrow \{ 1, 2, \cdots k \}$ which takes $a  \in \{ 1. \cdots , k \}$ to $\tau (a) \in \{ 1, \cdots , k \} $ acts as follows on the tensor product states 
\bea 
\tau \left (  e_{ i_1} \otimes e_{ i_2} \otimes  \cdots \otimes e_{ i_k } \right ) 
= \left (  e_{ i_{ \tau (1)} } \otimes e_{ i_{ \tau (2) } } \otimes \cdots e_{ i_{ \tau  (k )} } \right ) 
\eea
On the tensor product vector space $V^{ \otimes k } $, the group element acts as $g^{ \otimes k } $ 
\bea 
&& g^{ \otimes k }  \left ( e_{ i_1 } \otimes e_{ i_2 } \otimes \cdots \otimes e_{ i_{k } }  \right )  =  g_{ i_1}^{ j_1} \cdots g_{ i_k}^{ j_k }  \left ( e_{ j_1 } \otimes e_{ j_2 } \otimes \cdots \otimes  e_{ j_{k } }  \right ) 
\eea
with the summation convention of repeated indices ($j_a$)  being understood. The trace is thus 
\bea 
&& \tr_{ V^{ \otimes k } } ( g^{ \otimes k } \tau  )  
= {\rm Coeff } \left (  e^{ i_1} \otimes e^{ i_2} \cdots \otimes  e^{ i_k } , g^{ \otimes k } \tau  \left (  e_{ i_1} \otimes e_{ i_2} \cdots \otimes e_{ i_k } \right ) \right )  \cr 
&& = {\rm Coeff } \left (  e^{ i_1} \otimes e^{ i_2} \cdots \otimes  e^{ i_k } , g^{ \otimes k } \left (  e_{ i_{ \tau (1)} } \otimes e_{ i_{ \tau (2) } } \otimes \cdots e_{ i_{ \tau  (k )} } \right ) \right )  \cr 
&& = {\rm Coeff } \left (  e^{ i_1} \otimes e^{ i_2} \cdots \otimes  e^{ i_k } , g_{ i_{ \tau  (1)}}^{ j_1} g_{ i_{ \tau  (2)}}^{ j_2}
\cdots  g_{ i_{ \tau  (k)}}^{ j_k }  \left ( e_{ j_1} \otimes e_{j_2} \otimes \cdots \otimes e_{ j_k} \right ) \right ) \cr 
&& = g_{ i_{ \tau (1)}}^{ i_1} g_{ i_{ \tau  (2)}}^{ i_2}
\cdots  g_{ i_{ \tau  (k)}}^{ i_k }  
\eea
For a permutation with cycle lengths $a \in \{ 1, \cdots , k \} $ with multiplicities $q_a$, the expression in the last line can be simplified as 
\bea 
 g_{ i_{ \tau (1)}}^{ i_1} g_{ i_{ \tau (2)}}^{ i_2}
\cdots  g_{ i_{ \tau  (k)}}^{ i_k }   = \prod_{ a=1}^k  ( \tr ( g^a ) )^{ q_a }  
\eea
This expresses the fact that each cycle of length $a$ contributes a factor of $ \tr (g^a)$ to the trace in $V^{ \otimes k }$. 
We thus conclude  that 
\bea 
\tr_{ V^{ \otimes k } } ( g^{ \otimes k } \tau )   = \prod_{ a=1}^k  ( \tr ( g^a ) )^{ q_a }  
\eea
Using this equation, the bosonic partition function in \eqref{bospartfun1}  can be expressed as  
\bea\label{bospartfun2}  
&& \cZ_{ \rm  sym}  ( G , V ; x  ) = { 1 \over |G|}  \sum_{ g \in G }  \sum_{ k =0}  { x^k \over   k! }   \sum_{ q \vdash k  } \left ( { k! \over \Sym ~ q } \right )   \tr_{ V^{ \otimes k } } ( g^{ \otimes k } \tau  ) 
\eea 
where we have replaced the sum over all permutations in $S_k$ in equation  \eqref{bospartfun1} with a sum over partitions $q$ of 
$\tr_{ V^{ \otimes k } } ( g^{ \otimes k } \tau  )  $ for a fixed $\tau $  with the cycle structure given by $q$.  The factor $  \left ( { k! \over \Sym ~ q } \right )  $ is the number of permutations with the cycle structure $q$. 
The equation \eqref{bospartfun2} can be manipulated further as follows 
\bea 
&&  Z_{\rm  sym}  ( G , V ; x  ) = { 1 \over |G|} \sum_{ g \in G }  \sum_{ k =0}^{ \infty }   x^k  \sum_{ q \vdash k  }  { 1 \over \prod_{ a=1}^k a^{ q_a} q_a! }  \prod_{ a =1}^k  ( \tr ( g^a ) )^{ q_a }  \cr 
&& = { 1 \over |G|}\sum_{ g \in G } \sum_{ k =0}^{ \infty }    \sum_{ q \vdash k  }  { x^{ \sum_{a} a q_a }  \over \prod_{ a=1}^k a^{ q_a} q_a! }  \prod_{ a =1}^k  ( \tr ( g^a ) )^{ q_a }  \cr 
&& = { 1 \over |G|}\sum_{ g \in G } \sum_{ q_1 , q_2 , \cdots  }  ~~~  \prod_{ a =1  }^{ \infty }   { 1 \over q_a! }   \left ( {  x^{ a } \over a }  \right )^{ q_a}  ( \tr ( g^a ) )^{ q_a } \cr 
&& = { 1 \over |G|}\sum_{ g \in G } \prod_{ a=1}^{ \infty} \sum_{ q_a =0 }^{ \infty }   { 1 \over q_a! }   \left ( {  x^{ a } \over a }  \right )^{ q_a}  ( \tr ( g^a ) )^{ q_a } \cr 
&& = { 1 \over |G|}\sum_{ g \in G } \prod_{ a=1}^{ \infty } \Exp \left [    { x^a \over a } \tr_V ( g^a )   \right ] \cr 
&& = { 1 \over |G|}\sum_{ g \in G } \Exp \left [ \tr_V \left (  \sum_{ a } { x^a \over a }  g^a \right ) )   \right ] \cr 
&& = { 1 \over |G|}\sum_{ g \in G } \Exp \left [ - \tr_V \left ( \log  ( 1 - x g ) \right ) \right ] \cr 
&& =  { 1 \over |G| }  \sum_{  g \in G } { 1 \over { \det  ( 1 -  x D^V ( g )  )  }}
\eea

We have thus derived the Molien-Weyl formula : 
\bea\label{MWformula} 
\cZ_{\rm sym}  ( G , V ; x  ) =  { 1 \over |G| }  \sum_{  g \in G } { 1 \over {\det  }  ( 1 -  x D^V ( g )  )  } 
\eea

\subsubsection{Projecting to a general representation $R $ of $G$ and fourier transform to constrained holonomy   } 

In the above, we have derived a formula, as a sum over group elements, for the generating function of multiplicities of the trivial  representation of $G$ in the bosonic Fock space of states generated by oscillators in the representation $V$ of $G$. The same steps allow the derivation of a formula for the generating function of the multiplicities of a general irreducible  representation $R $ of $G$ in the Fock space.
This generating function, denoted $\cZ_{\rm sym ; R }  ( G , V ; x  )$, can be written as a sum over the degree $k$ of the oscillators as 
\bea 
\cZ_{ {\rm sym} ; R }  ( G , V ; x  ) = \sum_{ k =0}^{ \infty } x^k \tr_{ \Sym^k ( V ) }(  P_{R  }  ) 
\eea
The trace is over the $k$-fold symmetrised tensor power of $V$ of the projector $P_{ R}$ for the irreducible representation $R $.  The projector is a sum over group elements : 
\bea 
P_{ R } = { d_{ R  } \over |G| } \sum_{ g \in G } \chi^{ R } ( g^{-1} )~ g 
\eea
i.e. an element of the group algebra, where $d_{ R } $ is the dimension of the irreducible representation and $\chi^{ R  } ( g ) $ is the character of the group  element $g$ in the representation $R  $. Inserting this projector in place of $P_0$ at the start of \eqref{bospartfun} and following the steps of the derivation 
 leads to 
 \bea 
 \cZ_{{\rm sym} ; R }  ( G , V ; x  ) =  { d_{ R }  \over |G| } 
  \sum_{  g \in G } { \chi^{ R } ( g^{-1} )  \over {\det  }  ( 1 -  x D^V ( g )  )  } 
 \eea 
 The determinant is a function of the conjugacy class of $g$, so the right hand side can be written as a sum over conjugacy classes. We can also perform a Fourier transform of $ \cZ_{\rm sym ; R }  ( G , V ; x  )$ to  obtain a class function $\cZ_{\rm sym ; C }  ( G , V ; x  )$
 \bea 
 \cZ_{ { \rm sym }  ; C  }  ( G , V ; x  ) &  = & \sum_{ R  } { \chi^{R} ( g_{\scaleto{C}{4pt}}  ) \over d_R }    \cZ_{{ \rm sym} ; R }  ( G , V ; x  ) \cr 
 &= & { 1 \over {\det  }  ( 1 -  x D^V ( g_{\scaleto{C}{4pt}})  )  }
 \eea 
 where $g_{\scaleto{C}{4pt}}$   is a group element in the specified  conjugacy class $C$.  We have used the character orthogonality relation 
 \bea
 \sum_{ R } \chi^{ R } ( g_{\scaleto{C}{4pt}}  ) \chi^{ R } ( g^{-1}  ) = 
 {  |G| \over |C|  } \delta ( \rm {  Class}  ( g ) , C  ) 
 \eea
where $|C|$  is the size of that conjugacy class $C$, to 
 to do the sum over $R $ and to constrain the sum over $g$ in \eqref{MWformula} 
 to a fixed conjugacy class.

We can apply these observations to the case of $G = S_N$.  The generating function for multiplicities of the irrep $R $ specialised to this case is renamed for simplicity as  
\bea 
\cZ_{ {\rm sym} ; R }  ( G = S_N , V = V_{ N }  ; x  ) \rightarrow  \cZ ( N , R  , x ) 
\eea 
We have, from the above observations, 
\bea 
\cZ ( N , R , x )  = \sum_{ p } { \chi^{ R } ( g_p^{-1} ) \over \Sym ~ p }  \cZ ( N , p , x ) 
\eea
where $\cZ ( n , p , x ) $ is given by the product formula \eqref{MainProp}. The Fourier-transformed, conjugacy class-labelled partition functions are recognised as none other than the  $\cZ ( n , p , x ) $ itself  : 
\bea 
 \cZ_{{\rm sym }; C  }  ( G , V = V_N  ; x  )  \rightarrow \cZ ( N , p , x ) 
\eea
In the path-integral formulation developed in \cite{GPIMQM-PI} the constraint of the sum over $g$ in \eqref{MWformula} to a fixed conjugacy class can be implemented as a constraint on the product of group elements associated with the links in the discretized thermal circle. This shows that $\cZ ( n , p , x ) $ as given by the product formula \eqref{MainProp} has an interpretation as a path integral with holonomy around the thermal circle constrained to a fixed conjugacy class.

 \subsection{ Molien-Weyl determinant weighted counting of bosons in multiple representations  }\label{MWderWghtBos} 
 
There is a natural generalisation of \eqref{MWformula} when we consider a Fock space generated by  bosonic 
oscillators $A_{ 1; m_1 }^{ \dagger} , A^{ \dagger}_{ 2 ; m_2 } , \cdots , A^{ \dagger}_{ s ; m_s}  $ transforming in a representation $V_1, V_2 \cdots , V_s $ with $ m_a  $ ranging over a basis for irrep $V_l$, the index $m$ ranges over $\dim V_l$ values.  We allow distinct parameters $x_1 , x_2 , \cdots , x_s$ weighting the different representations: 
\bea 
\cZ_{\rm  sym}  ( G ; V_1 , V_2 \cdots , V_s  ; x_1 , x_2 , \cdots , x_s  ) =  { 1 \over |G| }  \sum_{  g \in G } \prod_{ l =1}^{ s} { 1 \over {  \det }  ( 1 -  x_l D^{V_l } ( g )  )  } 
\eea
We prove this for the case of two representations. The extension to the general $s$ case proceeds through an evident generalisation. 

\bea 
&& \cZ_{\rm  sym}  ( G ; V_1 , V_2 ; x_1 , x_2 ) 
={ 1 \over |G| }  \sum_{ g \in G }  \sum_{ k_1 =0}^{ \infty } \sum_{ k_2 =0}^{ \infty } 
x_1^{k_1} x_2^{ k_2} 
tr_{ \Sym^{ k_1} ( V_1 )  \otimes \Sym^{ k_2} ( V_2 ) }  ( g ) \cr 
&& = { 1 \over |G| }  \sum_{ g \in G } \sum_{ k_1 =0}^{ \infty } \sum_{ k_2 =0}^{ \infty }  { x_1^{ k_1} x_2^{ k_2}  \over k_1 ! k_2! } \sum_{ \tau_1 \in S_{ k_1} }\sum_{ \tau_2 \in S_{ k_2} }
\tr_{ V_1^{ \otimes k_1} } ( g^{ \otimes k_1} \tau_1 ) \tr_{ V_2^{ \otimes k_2} } ( g^{ \otimes k_2} \tau_2 ) \cr 
&& = { 1\over |G| }  \sum_{ g \in G } \sum_{ k_1, k_2} \sum_{ p \vdash k_1 } \sum_{ q \vdash k_2 }  {  x_1^{ k_1} x_2^{ k_2} \over \Sym ~p ~~ \Sym ~ q } 
\prod_{ a } \tr_{ V_1} ( g^a )^{ p_a}  \prod_{ b } \tr_{ V_2} ( g^b )^{ q_b } \cr 
&& =  { 1\over |G| }  \sum_{ g \in G } \sum_{ p_1 , p_2 \cdots } \sum_{ q_1, q_2 \cdots } 
{ x_1^{ \sum_a a p_a} x_2^{ \sum_b b q_b } \over \prod_a a^{ p_a} p_a! \prod_b b^{ q_b } q_b! } \prod_{  a } \tr_{ V_1} ( g^a )^{ p_a}  \prod_{ b } \tr_{ V_2} ( g^b )^{ q_b } \cr 
&& =  { 1\over |G| }  \sum_{ g \in G } \sum_{ p_1 , p_2 \cdots } \sum_{ q_1, q_2 \cdots } 
   \prod_{ a }  {1 \over p_a! }  ({x_1^a \tr_{ V_1} ( g^a ) \over a} ) ^{ p_a}    \prod_{ b }  {1 \over q_b! }  ({x_2^b \tr_{ V_2} ( g^b ) \over b} ) ^{ q_b } \cr 
   &&  =  { 1\over |G| }  \sum_{ g \in G } \prod_{ a } \sum_{ p_a }  {1 \over p_a! }  ({x_1^a \tr_{ V_1} ( g^a ) \over a} ) ^{ p_a}    \prod_{ b } \sum_{ q_b } 
    {1 \over q_b! }  ({x_2^b \tr_{ V_2} ( g^b ) \over b} ) ^{ q_b } \cr 
&& =  { 1\over |G| }  \sum_{ g \in G } \prod_a \exp ( { x_1^a\over a}  \tr_{ V_1} ( g^{ a } ) )\prod_b \exp ( { x_1^b \over b}  \tr_{ V_2} ( g^{ b } ) ) \cr 
&& = { 1\over |G| }  \sum_{ g \in G }  \exp ( \tr_{ V_1} \sum_{ a } { x_1^a D^{ V_1} ( g^a ) \over a } ) 
                                        \exp ( \tr_{ V_2} \sum_{ b } { x_2^b D^{ V_2} ( g^b  ) \over b  } ) \cr 
 && = { 1\over |G| }  \sum_{ g \in G }  \exp (  - \tr_{ V_1 } \log ( 1 - x_1 D^{ V_1} ( g ) ) ) 
                                        \exp (  - \tr_{ V_2 } \log ( 1 - x_2 D^{ V_2} ( g ) ) ) \cr 
 && = { 1\over |G| }  \sum_{ g \in G }  { 1 \over \det ( 1 - x_1 D^{ V_1} ( g ) ) }  { 1 \over \det ( 1 - x_2 D^{ V_2 } ( g ) ) } 
\eea
It is useful to define 
\bea\label{defzva}  
Z_{ V }^{ (g) }  ( x ) = { 1 \over \det ( 1 - x D^{ V} ( g ) ) } 
\eea
which allows us to  write 
\bea 
&& \cZ_{\rm  sym}  ( G ; V_1 , V_2 ; x_1 , x_2 )  = { 1\over |G| }  \sum_{ g \in G }Z_{ V_1  }^{ (g) }  ( x_1  ) Z_{ V_2  }^{ (g) }  ( x_2  ) 
\eea
and more generally 
\bea\label{resultdirectsum}  
\cZ_{\rm  sym}  ( G ; V_1 , V_2 \cdots , V_s  ; x_1 , x_2 , \cdots , x_s  ) =  { 1 \over |G| }  \sum_{  g \in G } \prod_{ a =1}^{ s} Z_{ V_a  }^{ (g) }  ( x_a   ) 
\eea

\subsection{ Molien-Weyl determinant formula derivation : Fermionic case }\label{MWderFerm} 
 
In the fermionic case, the $k$-oscillator state space is the antisymmetrized tensor product $\Lambda^k ( V )$. The generating function of 
fermionic states which are $G$-invariant is given by 
\bea 
\cZ_{\hbox{\tiny ferm}  } ( G ,  V ; x ) & = &  \sum_{ k =0 }^{ \infty } { x^k }  \tr_{ \Lambda^k ( V ) } P_0 
\eea 
where the projector to $G$-invariants is 
\bea 
P_0 = {1\over |G|} \sum_{ g \in G } g 
\eea
The trace over the $k$-fold anti-symmetrized tensor space $\Lambda^k ( V ) $  can be written as trace over the $k$-fold  tensor power $V^{ \otimes k } $  along with 
an anti-symmetrization projector constructed as a sum of permutations $\tau $  in $S_k$, weighted by the sign of the permutation which is denoted 
$(-1)^{ \tau} $ 
\bea 
Z_{\hbox{\tiny ferm}} ( G ,  V ; x )  && = \sum_{ k=0 }^{ \infty } \sum_{ \tau \in S_k }  x^k { (-1)^{ \tau } \over k ! } \tr_{ V^{ \otimes n } } ( P_0 \tau ) 
\eea 
We convert the sum over permutations in $S_k$ to a sum over conjugacy classes $q$  in $S_k$, specified as $ q = [ i^{ q_i} ] $ as in the bosonic case. The sign of a permutation $\tau $ in class $q$ is a class function 
which can be expressed as 
\bea 
(-1)^{ \tau } = (-1)^{ \tau^{(q)} } = (-1)^{ \sum_{i}  ( i -1 ) q_i}  
\eea 
This expresses the fact that odd-length cycles are even sign permutations while even-length cycles are odd-sign permutations.
The  fermionic partition function is thus 
\bea  Z_{ \hbox{\tiny ferm}  } ( G ,  V ; x ) 
&& = \sum_{ k =0}^{ \infty } \sum_{ q \vdash k }  { ( -1 )^{ \tau^{(q) } }\over \Sym ~ q } \tr_{ V^{ \otimes n } } ( \tau^{ (q)} P_0) ) \cr 
&& = \sum_{ k =0}^{\infty } \sum_{ q \vdash k } \sum_{ g } { 1 \over |G| }  { ( -1 )^{ \tau (q) } \over \Sym ~ q } \tr_{ V^{ \otimes k } } ( g^{ \otimes k } \tau^{ (q)} ) \cr 
&& =\sum_{ g } { 1 \over |G|  } \sum_{ k =0}^{ \infty }  x^k \sum_{ q \vdash k } \prod_{ a =1}^{ k } { (-1)^{ (a-1) q_a } \over a^{ q_a} q_a! }  ( \tr g^a )^{ q_a}  \cr 
&& = \sum_{ g } { 1 \over |G|  } \sum_{ k =0}^{ \infty }   \sum_{ q \vdash k } \prod_{ a =1}^{ k } { (-1)^{ (a-1) q_a } x^{ a q_a} \over a^{ q_a} q_a! }  ( \tr g^a )^{ q_a}  \cr 
&& = \sum_{ g } { 1 \over |G| } \sum_{ q_1 , q_2 \cdots } \prod_{ a } {  (-1)^{ q_a } \over q_a ! }  \left( { (-x )^a \tr g^a \over a } \right )^{ q_a} \cr 
&& = \sum_{ g } { 1 \over |G| } \prod_{ a } \sum_{ q_a =0 }^{ \infty } 
  {  (-1)^{ q_a } \over q_a ! }  \left ( { (-x )^a \tr g^a \over a } \right)^{ q_a } \cr
&& = \sum_{ g } { 1 \over |G| } \prod_{ a =1 }^{ \infty }   e^{   - { ( -x)^a \over a } \tr g^a }  \cr 
&& = \sum_{ g } { 1 \over |G| }   e^{   - \sum_{ a =1}^{ \infty }  { ( -x)^a \over a } \tr g^a  }\cr 
&& = \sum_{ g } { 1 \over |G| } e^{ \sum_{ a =1}^{ \infty } { ( -1)^{ a+1} x^a \tr g^a \over a } }\cr 
&& = \sum_{ g } { 1 \over |G| }e^{\tr  \sum_{ a =1}^{ \infty } { ( -1)^{ a+1} x^a D^V ( g^a ) \over a } } \cr 
&& = \sum_{ g }  { 1 \over |G| } e^{\tr D^V  \left (    x g - { x^2   g^2  \over 2 } + { x^3 g^3 \over 3 } + \cdots  \right )   } \cr 
&& =  \sum_{ g }  { 1 \over |G| } e^{\tr  \log ( 1 + x D^V ( g ) ) } \cr 
&& =\sum_{ g }  { 1 \over |G| }  \det ( 1 + x D^V ( g ) ) 
\eea
We have thus proved the Molien-Weyl formula for $G$-invariant fermionic state counting 
\bea 
Z_{ \hbox{\tiny ferm}} ( G ,  V ; x )  = \sum_{ g }  { 1 \over |G| }  \det ( 1 + x D^V ( g ) ) 
\eea

\section{ Gauged permutation invariant Matrix partition function using Molien-Weyl formula}\label{sec:MWtoLCM}  

We return to the specific case of interest at hand, the  matrix oscillators $A^{ \dagger}_{ij}$  transforming as $(V_N \otimes V_N)$ where $V_N$ is the natural representation of $S_N$ of  dimension $N$, with the decomposition into irreducible representations 
\bea 
V_{N } = ( V_0 \oplus  V_H )
\eea
where $V_0$ is the trivial one-dimensional representation and $V_H$ is the $(N-1)$ dimensional representation.  In this section, we will reproduce the formula \eqref{MainProp} for the generating function of bosonic states using the Molien-Weyl formula \eqref{MWformula}.  It is useful to recall that the natural representation of dimension $N$ can be described as 
\bea 
V_{ N } = \hbox{ Span } \{ e_i  ~~ | ~~ i \in \{ 1, 2, \cdots , n \} \} 
\eea
where permutations $\sigma $ act as  linear operators $D^{ V_N} ( \sigma )$ defined as
\bea\label{natact}  
D^{ V_N} ( \sigma )  : e_i \rightarrow e_{ \sigma^{-1} ( i ) } 
\eea
With the left-to-right multiplication convention 
\bea 
\sigma_1 \sigma_2 ( i ) = \sigma_2 ( \sigma_1(i)  ) 
\eea
the linear operators obey the homomorphism condition 
\bea 
D^{ V_N} ( \sigma_1 ) D^{ V_N } ( \sigma_2 ) = D^{ V_N } ( \sigma_1 \sigma_2 )  
\eea

For a cyclic permutation $ \sigma_a =  ( 1, 2, \cdots , a )$ acting on the vector space $V_{ a} $ of dimension $a$, using the action in \eqref{natact}. 
The eigenvalues of this action are multiples of $\omega_a = e^{ 2 \pi i \over a } $ : 
\bea\label{specZa} 
\omega_a^{ t } \qquad \qquad t \in \{ 0 , 2, \cdots , a-1   \} 
\eea
$\sigma_a$ generates the cyclic group $\mZ_a$ and $V_a$ is isomorphic to the regular representation of $\mZ_a$.  The corresponding eigenvectors are 
\bea 
| \omega_a^t \rangle  = e_1  + \omega_a^{ -t} e_2 + \cdots + \omega_a^{ - (a-1)t  } e_a 
\eea

As in the derivation of \eqref{MainProp} in section \ref{FirstDerBosPart} it is convenient to consider, in the first instance, permutations $\sigma \in S_N $ in a conjugacy class $p =  [ a_1^{ p_1} , a_2^{  p_2} ] $, i.e. $\sigma $ has $p_1$ cycles of length $a_1$ and $p_2 $ cycles of length $a_2$.  The permutation $\sigma $ belongs to a subgroup of the form  of a product of cyclic groups:  $\mZ_{ a_1}^{ \times p_1} \times \mZ_{ a_2}^{\times  p_2} \subset S_N$. We can therefore use the spectrum 
\eqref{specZa} to deduce  that the  eigenvalues of the matrix $D^{ V_N } ( \sigma )$ are :
\bea 
 \omega_{a_1}^{ t_1 } ,  t_1 \in \{ 0 , \cdots ,  a_1 -1    \}   ~~~ : ~~~ &&\hbox{ Multiplicity } ~~ = ~~  p_1\cr 
 \omega_{a_2}^{ t_2  } , t_2  \in \{ 0  , \cdots ,  a_2 -1    \} ~~~ : ~~~ && \hbox { Multiplicity } ~~~ = ~~  p_2 
 \eea
 In the tensor product $V_N \otimes V_N$, the eigenvectors are of two types, as follows : 
 \bea\label{eigenvaluepairs} 
&& \hskip1cm \hbox{ Same eigenvalue in both tensor factors } \cr 
&&  | \omega_{a_1}^{ t_1} ; \alpha_1 \rangle \otimes | \omega_{a_1}^{ t_2} , \alpha_2 \rangle ~~~~ \alpha_1 \in \{ 1 \cdots p_1 \} ~~ \alpha_2 \in \{ 1 \cdots p_1  \}  \cr 
 && | \omega_{a_2}^{ t_1} ; \alpha_1 \rangle \otimes | \omega_{a_2}^{ t_2} , \alpha_2 \rangle ~~~~ \alpha_1 \in \{ 1 \cdots p_2 \} ~~ \alpha_2 \in \{ 1 \cdots p_2 \} \cr 
&& \hskip1cm \hbox{ Different  eigenvalues in the two tensor factors } \cr 
 && | \omega_{a_1}^{ t_1} ; \alpha_1 \rangle \otimes | \omega_{a_2}^{ t_2} , \alpha_2 \rangle ~~~~ \alpha_1 \in \{ 1 \cdots p_1 \} ~~ \alpha_2 \in \{ 1 \cdots p_2 \} \cr   
 && | \omega_{a_2}^{ t_2 } ; \alpha_2 \rangle \otimes | \omega_{a_1}^{ t_1} , \alpha_1 \rangle ~~~~ \alpha_1 \in \{ 1 \cdots p_2 \} ~~ \alpha_2 \in \{ 1 \cdots p_1 \}
 \eea 
The index $\alpha_1 $ identifies one of the $Z_{ a_1} $ in the product $\mZ_{ a_1}^{\times  p_1} $ while the index $\alpha_2$ identifies one of  the $\mZ_{   a_2}$ factors in the product $\mZ_{ a_2}^{ \times p_2} $.  This leads to 
 \bea\label{prodDet}
 { \det }  ( 1 - x D^{ V_N \otimes V_N} ( g ) )  = 
 \prod_{ t_1 , t_2 =0}^{ a_1 -1 }  ( 1 - x \omega_{a_1}^{ t_1 + t_2 } )^{ p_1^2 }  \prod_{ t_1, t_2  =0 }^{ a_2-1  }  ( 1 - x \omega_{a_2}^{ t_1 + t_2 } )^{ p_2^2 }  \prod_{ t_1 =0 }^{ a_1 -1  } \prod_{ t_2 = 0 }^{ a_2 -1  } ( 1 - x \omega_{a_1}^{ t_1 } \omega_{a_2}^{ t_2} )^{ 2 p_1 p_2 }  \cr 
 && 
 \eea
 A useful lemma is : \\
\noindent 
{\bf Lemma 1: } 
 \begin{equation}\label{prodroots} 
 \boxed{ ~~~~~
\prod_{ t =0}^{ a-1 }  ( 1 - X \omega_a^{ t } ) = ( 1 - X^{ a } ) ~~~~~
} 
 \end{equation} 
The proof follows by observing that the coefficient of $X^{ c } $ when the product is expanded out is : 
\bea  
(-1)^c  \sum_{ t=0  }^{ a-1 }  \omega_{a}^{ c t } && = { ( 1 - \omega_{a}^{ c a} )  \over ( 1- \omega_a^c ) }  \hbox{ for } c \in \{ 1 , 2, \cdots , a -2  \}  \cr 
&& =1\hbox{   for }  c \in \{ 0 , a-1  \} 
\eea
 This implies 
 \bea\label{sameeig1}  
  \prod_{ t_1 , t_2 =0}^{ a_1 -1  }  ( 1 - x \omega_{a_1}^{ t_1 + t_2 } ) = \prod_{ t_1 =0}^{ a_1 -1  } ( 1 - x^{ a_1}  \omega_{a_1}^{ a_1  t_1 } )  = \prod_{ t_1 =0}^{ a_1 -1  } ( 1 - x^{ a_1}  ) = ( 1- x^{ a_1} )^{ a_1} 
 \eea
 Similarly 
 \bea\label{sameeig2} 
   \prod_{ t_1 , t_2 =0 }^{ a_2 -1 }  ( 1 - x \omega_{a_2}^{ t_1 + t_2 } ) = \prod_{ t_1 =0 }^{ a_2 -1 } ( 1 - x^{ a_2}  \omega_{a_2}^{ a_2  t_1 } )  = \prod_{ t_1 =0}^{ a_2 -1  } ( 1 - x^{ a_2}  ) = ( 1- x^{ a_2} )^{ a_2} 
 \eea
To calculate the mixed terms, we will need a second lemma, which we will call the roots and LCM Lemma since it connects the roots of unity eigenvalues and the Least common multiples : \\
\noindent 
{\bf Lemma 2:  Roots-and-LCM-Lemma  } \\
\begin{equation}\label{RLCML} 
\boxed{ ~~~~
\prod_{ t_1 =0}^{ a_1 -1 } ( 1 - x^{ a_2}  \omega_{a_1}^{ a_2 t_1  }  ) 
   = ( 1 - x^{ L ( a_1 , a_2 )  } )^{ a_1 a_2 \over  L (  a_1 , a_2 ) } = ( 1 - x^{ L ( a_1 , a_2 )  } )^{ G (  a_1 , a_2 ) } 
~~~~ } 
\end{equation} 
We will prove the Lemma shortly, but using the Lemma we can now calculate  the  mixed terms in  \eqref{prodDet}. We note that  
\bea
 &&  \prod_{ t_1 =0}^{ a_1-1  } \prod_{ t_2 = 0 }^{ a_2 -1  } ( 1 - x \omega_{a_1}^{ t_1 } \omega_{a_2}^{ t_2} )
  =   \prod_{ t_1 =0}^{ a_1-1  } ( 1 - x^{ a_2}  \omega_{a_1}^{ a_2 t_1  }  ) 
   = ( 1 - x^{ L ( a_1 , a_2 )  } )^{ a_1 a_2 \over  L (  a_1 , a_2 ) }  = ( 1 - x^{ L ( a_1 , a_2 ) } )^{ G (a_1 , a_2 ) } \cr 
   &&  ~~
 \eea
which gives 
\bea 
  \prod_{ t_1 =0}^{ a_1 -1 } \prod_{ t_2 = 0 }^{ a_2 -1 } ( 1 - x \omega_{a_1}^{ t_1 } \omega_{a_2}^{ t_2} )^{ 2 p_1 p_2 } 
=  ( 1 - x^{ L ( a_1 , a_2 ) } )^{2 p_1 p_2  G (a_1 , a_2 ) } 
\eea
 Putting these products together with their $p$-dependent multiplicities as given in \eqref{eigenvaluepairs}, we have 
 \bea 
  { \det }  ( 1 - x D^{ V_N \otimes V_N} ( g ) )  
  = ( 1 - x^{ a_1 } )^{ a_1 p_1^2 } ( 1 - x^{ a_2} )^{ a_2 p_2^2 } ( 1 - x^{ L ( a_1 , a_2 ) } )^{ 2 p_1 p_2 G ( a_1 , a_2 ) } 
 \eea
In other words, for $ [ g ] = [ a_1^{  p_1 } , a_2^{ p_2 } ] $. 
\bea 
{ 1 \over   { \det }  ( 1 - x D^{ V_N \otimes V_N} ( g ) )   } 
= { 1 \over ( 1 - x^{ a_1 } )^{ a_1 p_1^2 } ( 1 - x^{ a_2} )^{ a_2 p_2^2 } ( 1 - x^{ L ( a_1 , a_2 ) } )^{ 2p_1 p_2  G ( a_1 , a_2)    } } 
\eea

\noindent 
{\bf  Proof of  the Roots-and-LCM-Lemma:  } \\
\vskip.2cm 
For the  greatest common divisor  of $a_1, a_2$, previously denoted $G ( a_1 , a_2 ) $ we will write for simplicity $G (a_1, a_2 ) =  h$. We have 
\bea 
&& a_1 =  h~ \hat a_1  \cr 
 && a_2 = h ~ \hat a_2 \cr 
 && G  ( \hat a_1 , \hat a_2 ) = 1 
\eea
for positive integers $h , \hat a_1 , \hat a_2 $. In the product over $ t_1 \in \{ 0, 2, \cdots , a_1-1  \}$ in \eqref{RLCML} 
it  is useful to write $t_1 =  q \hat a_1 + s $, with $ s \in  \{ 0 , \cdots , \hat a_1 -1  \}  , q \in \{ 0, \cdots , h -1    \}   $. So we have  to calculate 
\bea 
\prod_{ t_1 =0 }^{ a_1 -1  } ( 1 - x^{ a_2}  \omega_{a_1}^{ a_2 t_1  }  )   = \prod_{ q=0}^{ h-1} \prod_{ s=0 }^{ \hat a_1 - 1  } ( 1 - x^{ a_2}  \omega_{a_1}^{ a_2 t_1  }  )  
\eea  
Note that 
\bea 
\omega_{a_1}^{ a_2 t_1  } = e^{ 2 \pi i a_2 t_1 \over a_1 } = e^{ 2 \pi i \hat a_2 t_1  \over \hat a_1 } 
= e^{ 2 \pi i \hat a_2 ( q \hat a_1 + s )   \over \hat a_1 } = e^{ 2 \pi i \hat a_2 s \over \hat a_1 } 
\eea
Since $G ( \hat a_1 , \hat a_2 ) =1 $, $e^{ 2 \pi i \hat a_2 \over \hat a_1 }  $ is a primitive root of unity of order $ \hat a_1$. 
This means, using Lemma \eqref{prodroots} 
\bea 
\prod_{ s =0  }^{ \hat a_1 -1  } ( 1 - X  e^{ 2 \pi i \hat a_2 s \over \hat a_1 } ) = ( 1 - X^{ \hat a_1 } ) 
\eea
We can write 
\bea 
\prod_{ t_1 =0}^{ a_1-1  } ( 1 - x^{ a_2}  \omega_{a_1}^{ a_2 t_1  }  )  
= \prod_{ q = 0}^{ h -1 }  \prod_{ s =0}^{ \hat a_1 -1 } ( 1 - x^{ a_2}  e^{ 2 \pi i s \hat a_2 \over \hat a_1 }  )  = \prod_{ q=0}^{ h-1} ( 1 - x^{ a_2  \hat a_1 } ) = ( 1 - x^{ a_2  \hat a_1 } )^h 
\eea
Now observe that 
\bea 
&& L ( a_1 , a_2 )  \equiv LCM ( a_1 , a_2 ) = LCM  ( h \hat a_1 , h \hat a_2 ) = h \hat a_1 \hat a_2 \cr 
&& a_1 a_2  = h^2 \hat a_1 \hat a_2 \cr 
&& h = {  a_1 a_2 \over LCM ( a_1 , a_2 ) }  
\eea
We conclude that 
\bea\label{distinctpairs}  
&& \prod_{ t_1 =0 }^{ a_1 -1  } ( 1 - x^{ a_2}  \omega_{a_1}^{ a_2 r_1  }  )  = 
 ( 1 - x^{ a_2  \hat a_1 } )^h  =  ( 1 - x^{ h \hat a_2  \hat a_1 } )^h  \cr 
&&  = ( 1 - x^{ L ( a_1 , a_2 ) }  )^{  { a_1 a_2 \over L ( a_1 , a_2 ) }  }  =  ( 1 - x^{ L ( a_1 , a_2 ) }  )^{  {G (  a_1 ,  a_2)    }  } 
\eea
For a general partitition $ [ a_1^{ p_1} , a_2^{ p_2} , \cdots , a_K^{ p_K} ] $, the tensor product states can be organised as in \eqref{eigenvaluepairs} 
according to whether the eigenvalues in the two factors of the tensor product $V_N \otimes V_N $ are the same or different.  When the eigenvalues are the same we get, collecting factors of the form in \eqref{sameeig1} and \eqref{sameeig2}, a factor in the determinant 
$ { \det }  ( 1 - x D^{ V_N \otimes V_N} ( g ) )  $ 
\bea 
\prod_{ i =0 }^{ a_i -1 }  ( 1- x^{ a_i } )^{ a_i } 
\eea
Collecting contributions to the determinant from cases where the eigenvalues are distinct, noting that the distinct pairs can be any $a_i , a_j $ with $i , j \in \{ 1, \cdots  , K \}$  and using the \eqref{distinctpairs} formula, we have 
\bea 
\prod_{ i  < j \in \{ 1, \cdots , K \} }   ( 1- x^{ L ( a_i , a_j ) })^{ G ( a_i , a_j ) } 
\eea
Collecting both cases and using the multiplicities of eigenvalues for the two cases from \eqref{eigenvaluepairs}  we obtain 
\bea 
\prod_{ i =1 }^{ K  }  ( 1- x^{ a_i } )^{ a_i p_i^2  } \prod_{ i  < j \in \{ 1, \cdots , K \} }   ( 1- x^{ L ( a_i , a_j ) })^{ 2 p_i p_j G ( a_i , a_j ) } 
\eea
which agrees with $\cZ ( N , p , x ) $. 

\section{ The refined partition function for general gauged $11$-parameter matrix  harmonic oscillator  system } 

In this section we derive the partition function for the general gauged permutation invariant matrix harmonic oscillator. The general permutation invariant quadratic function of matrix variables  was described in \cite{LMT} and  the representation theoretic diagonalisation was given in \cite{PIGMM}. The canonical quantization of the matrix quantum system was described  and the canonical partition function of the full model, including permutation invariant and non-invariant sectors was given in \cite{PIMQM}. The Molien-Weyl formula, which was recovered physically as a continuum limit of the lattice formulation of the gauge invariant theory in \cite{GPIMQM-PI}, turns out, as we show here, to be a powerful tool for calculating the partition function of the general model. 

The partition function of the  $11$-parameter harmonic oscillator system, with $S_N$ invariance imposed as a gauge symmetry,  takes the form 
\bea 
&& \cZ ( N ,\beta_{0}^{ ab }  , \beta_{H }^{ ab } , \beta_{ 2}   , \beta_{ 3 }  )  = \cr 
 &&   \tr_{ \cH } \cP_0  \exp \left [ { - \sum_{ a, b =1}^2 \beta_{0}^{ ab } A_{0; a}^{\dagger } A_{0; b} -  \sum_{ a, b =1}^3 \beta_{H }^{ ab }  \sum_{ m=1}^{ \dim V_H} A_{H, a; m }^{\dagger }A_{H, b ; m } - \beta_{ 2} \sum_{ m =1}^{ \dim V_2 }  A_{2 ; m }^{\dagger }A_{2 ; m  }  -  \beta_{ 3 } \sum_{ m =1}^{ \dim V_3 }  A_{3 ; m }^{\dagger }A_{3 ; m } }\right ] \cr 
&& 
\eea
The  path integral formulation for the quantum system, as a limit of a lattice formulation where the covariant derivative is described in terms of parallel transport using group elements, is given in \cite{GPIMQM-PI}. The Hilbert space $\cH$ is the Fock space generated by the matrix oscillators $A^{\dagger}_{ ij}$. The exponent uses linear combinations of the $A^{ \dagger}_{ ij}$ oscillators organised according to the decomposition of  the matrix oscillators into irreducible representations of $S_N$ : 
\bea\label{VNVNdecomp} 
V_N \otimes V_N = 2 V_0 \oplus 3 V_H \oplus V_2 \oplus V_3 
\eea
where $V_0$ is the one-dimensional representation of $S_N$, $V_H$ is the $(N-1)$ dimensional hook representation corresponding to the Young diagram with row lengths $[N-1,1]$, $V_2 $ is the representation of dimension $ N ( N -3)/2$ associated with Young diagram $[N-2,2] $, while $V_3$ is the irrep of dimension $ (N-1)(N-2)/2$ associated with the Young diagram $[N-2,1,1]$.  The creation operators 
\bea 
A_{0; a}^{\dagger } ~~~ , ~~~ A_{H,a  ; m }^{ \dagger}  ~~~ , ~~~  A_{2  ; m }^{ \dagger}   ~~~~ , ~~~ A_{3  ; m }^{\dagger} 
\eea
transform in the four irreps, respectively, $V_0 , V_H , V_2 , V_3$ as in \eqref{VNVNdecomp}. The $a$ indices run over the multiplicity spaces and these creation operators can be expressed as linear combinations of $A^{ \dagger}_{ ij}$ using the Clebsch-Gordan coefficients associated with \eqref{VNVNdecomp}, as explained in more detail in \cite{PIMQM}. Similar remarks apply to the annihilation operators. The parameters $\beta_{ 0}^{ ab} $ form a symmetric $2\times 2 $ matrix (with $3$ parameters), the $\beta_{ H}^{ ab}$ form a symmetric $3 \times 3 $ matrix, so that alongside $\beta_2, \beta_3$ there is a the total number of parameters is $ 3 +6 + 1 +1 = 11$. By diagonalising the matrices $\beta_0^{ Hab}$ and $\beta_{ H}^{ ab}$, we find that the thermal partition function of the invariant sector depends on $ 2 +3 + 1 + 1 = 7$ parameters 
(as does the partition function of the full Hilbert space computed in 
\cite{PIMQM}). We thus find that the most general  partition function takes the form 
\bea
&&   \cZ ( N , \beta_0^a  , \beta_H^a , \beta_2 , \beta_3 ) \cr 
&&  = 
\tr_{ \widetilde \cH }  \cP_0 \exp \left [ { -  \sum_{ a =1}^2 \beta_{0}^{ a } \tilde A_{0; a}^{\dagger } \tilde A_{0;a} - \sum_{ a =1}^3 \beta_{H }^{ a }  \sum_{ m=1}^{ \dim V_H} \tilde A_{H, a; m }^{\dagger } \tilde A_{H, b ; m } - \beta_{ 2} \sum_{ m =1}^{ \dim V_2 }  \tilde A_{2 ; m }^{\dagger } \tilde A_{2 ; m  }  -  \beta_{ 3 } \sum_{ m =1}^{ \dim V_3 }  \tilde A_{3 ; m }^{\dagger } \tilde A_{3 ; m } }\right ]  \cr 
&& 
\eea 
where $\beta_0^a , \beta_H^a$ are the eigenvalues of the matrices $ \beta_0^{ ab} , \beta_H^{ ab}$ respectively. The Hilbert space 
$\widetilde \cH$ is the Fock space generated by the $7$ independent creation operators $ \tilde A_{0; a}^{\dagger } , \tilde A_{H; a}^{\dagger }  , \tilde A_2^{ \dagger} , \tilde A_3^{ \dagger}$. 


The $S_N$ invariant partition function is a weighted counting of Fock space states generated by the seven types of oscillators 
\bea 
&&  \tilde A_{0, a}^{\dagger } , a \in \{ 1, 2 \}  ~~\hbox{ weighted by } ~~ e^{ - \beta_{0}^{ a }  } \cr 
 && \tilde A_{H, a; m }^{\dagger } , a \in \{ 1, 2 , 3  \} ~~ \hbox{ weighted by } ~~ e^{ - \beta_{0}^{ a }  }  \cr 
 && \tilde A_{2 ; m }^{\dagger } ~~~ \hbox{ weighted by } ~~~ e^{ - \beta_2 } \cr 
 && \tilde A_{3 ; m }^{\dagger } ~~~  \hbox{ weighted by } ~~~ e^{ - \beta_3}  
 \eea
with the insertion of the $S_N$ projector. As explained in section \eqref{MWderWghtBos}, the weighted counting of invariants  takes the form 
\bea\label{proddetsref}  
 &&  \cZ  ( N , \beta_0^a  , \beta_H^a , \beta_2 , \beta_3 ) = 
  = { 1 \over N ! } \sum_{ \sigma }   \left ( \prod_{a=1}^2  Z^{ (\sigma) }_{ V_0 } ( e^{ - \beta_0^a} ) \right ) \left (  \prod_{a=1}^3  Z_{ V_H }^{ (\sigma )}  (e^{ - \beta_H^a} ) \right ) 
  Z_{ V_2}^{ (\sigma)}  ( e^{ - \beta_2}  )  Z_{ V_3 }^{( \sigma ) }  ( e^{ - \beta_3 } ) \cr 
  && 
\eea
where each factor is an inverse determinant of Molien-Weyl form ( see in particular equations \eqref{resultdirectsum}  and \eqref{defzva} ).

\subsection{ Molien-Weyl determinants for $V_N, V_0 , V_H$.  }\label{sec:V0VH} 

The natural representation $V_N$ decomposes as 
\bea 
V_N = V_0 \oplus V_H 
\eea
For the Fock space generated by an oscillator in the trivial one-dimensional representation, the 
counting function is 
\bea 
\cZ_{ V_0 } ( x ) =  \sum_{ p } { 1 \over \Sym ~ p } { 1 \over ( 1  - x D^{ V_0} ( \sigma^{(p)}  ) ) }  =
 \sum_{ p } { 1 \over \Sym ~ p } { 1 \over ( 1  - x   ) } 
\eea
Next we consider oscillators in the representation $V_N$ and a permutation $\sigma $ with cycle structure of the form $ [ a_1^{ p_1} , a_2^{ p_2} \cdots a_K^{ p_K}  ]$. 
This permutation belongs to a subgroup of the form $\mZ_{ a_1}^{ p_1} \times Z_{ a_2}^{ p_2} \cdots \times Z_{ a_K}^{ p_K}$ in $S_N$. The eigenvectors are of the form 
\bea 
| \omega_{ a_i}^{ t_i } ; \alpha_i \rangle
~~\hbox{ with } ~~~ i \in \{ 1, 2, \cdots , K \} ~~, ~~ t_i \in \{ 0 , 1, \cdots , a_i -1 \} ~~, ~~  \alpha_{ i  } \in \{ 1, 2, \cdots , p_i \}
\eea
with eigenvalues $ \omega_{ a_i}^{ r_i } $ each with multiplicity $p_i$. The total number of eigenvectors is 
$p_1 a_1  + p_2 a_2  + \cdots + p_K a_K =N$ and these eigenvectors span $V_N$. Therefore, the determinant is (using \eqref{prodroots})
\bea\label{MWVN}  
\det (  1 - x D^{V_N} ( \sigma ) ) = \prod_{ i=1}^K \prod_{ t_i =0}^{a_i-1 }   ( 1 - x \omega_{a_i}^{ t_i} )^{ p_i}  
= \prod_{ i=1}^K  {  ( 1 - x^{a_i}  )^{ p_i}  } 
\eea
We also have 
\bea 
\det (  1 - x D^{V_0} ( \sigma ) ) = ( 1 - x ) 
\eea
and 
\bea 
\det (  1 - x D^{V_N} ( \sigma ) ) = \det (  1 - x D^{V_0 \oplus V_H } ( \sigma ) ) = \det (  1 - x D^{V_0} ( \sigma ) ) \det (  1 - x D^{V_H} ( \sigma ) ) 
\eea
We conclude that 
\bea\label{MWVH}  
\det (  1 - x D^{V_H} ( \sigma ) )  = { 1 \over ( 1 - x ) } \prod_{ i =1}^K ( 1 - x^{ a_i} )^{ p_i} 
\eea

The Molien-Weyl formula then becomes : 
\bea 
\cZ_{ V_H }  (  x  ) = \sum_{ p \vdash N    }   { ( 1- x ) \over \Sym ~ p }  \prod_{ i } { 1 \over ( 1 -x^{ a_i} )^{  p_i } } 
\eea

\subsection{ Using $S^2 V_N $ and $\Lambda^2 V_N$ to get the determinants for $V_2,V_3$  }\label{sec:symmirreps} 

The eigenvectors of $\sigma $, with cycle structure $ [ a_1^{ p_1} , a_2^{ p_2} , \cdots , a_K^{ p_K} ] $ in $V_N \otimes V_N $ are 
\bea 
|\omega_{ a_i}^{ t_i} ; \alpha_i \rangle \otimes |\omega_{ a_j}^{ t_j} ; \alpha_j  \rangle 
\eea
where 
\bea 
&& i , j \in \{ 1, 2, \cdots , K \} \cr 
&& t_i \in \{ 0,1, \cdots , a_i-1 \}   \cr 
&& t_j \in \{ 0 , 1, \cdots , a_j - 1 \} \cr 
&& \alpha_i \in  \{ 0 , 1, \cdots , p_i  \}  ~~, ~~ \alpha_j \in \{  1, \cdots , p_j  \} 
\eea
It is known  that the symmetric tensor power decomposes as 
\bea 
S^2 ( V_H ) = V_0 \oplus V_H \oplus V_2 
\eea
and 
\bea 
\Lambda^2 V_H = V_3 
\eea
These decompositions are described in detail in \cite{PIGMFC}. A dimension count to check the first equation is  
\bea\label{decompS2N}  
 { N( N-1 ) \over 2 }  = 1 + ( N -1 ) + { N ( N -3 ) \over 2 } \cr 
\eea
One also checks that the hook formula for $ [ N-2,1,1]$ agrees with $ ( N -1) ( N -2) /2 $. 

Since $V_N = V_0 \oplus V_H$, we can use the above to find 
\bea 
S^2 ( V_N ) &&  = S^2 ( V_0 ) \oplus S^2 ( V_H ) \oplus \Sym  \left ( (  V_H \otimes V_0 )  \oplus ( V_0 \otimes V_H ) \right )  \cr 
               && = V_0 \oplus ( V_0 \oplus V_H \oplus V_2 ) \oplus V_H \cr 
               && = 2 V_0 \oplus 2 V_H \oplus V_2 
\eea
and 
\bea\label{decompL2N}  
\Lambda^2 ( V_N )  = \Lambda^2 ( V_H ) \oplus \Lambda \left(  ( V_0 \otimes V_H)  \oplus ( V_0 \otimes V_H ) \right ) = V_3 \oplus  V_H 
\eea
These equations imply that 
\bea 
\det ( 1 - x D^{ S^2 (V_N ) } ( \sigma )  ) = \det ( 1 - x D^{ V_0 } ( \sigma ) )^2 \det ( 1 - x D^{ V_H } ( \sigma ) )^2 
\det ( 1 - x D^{ V_2 } ( \sigma ) ) 
\eea
and 
\bea 
\det ( 1 - x D^{ \Lambda^2 (V_N ) } ( \sigma )  )  = \det ( 1 - x D^{ V_H } ( \sigma ) ) \det ( 1 - x D^{ V_3 } ( \sigma ) ) 
\eea
which can be used to calculate $ \det ( 1 - x D^{ V_2 } ( \sigma ) )$ and   $  \det ( 1 - x D^{ V_3 } ( \sigma ) ) $ 
from the determinants of the symmetrised and anti-symmetrised powers along with those for $V_0, V_H$. The determinants for $V_0$ and $V_H$ are already given in section \ref{sec:V0VH}. We will now calculate the determinants for the symmetrised and anti-symmetrised square of $V_N$.

\subsection{ Molien-Weyl determinant for  $S^2(V_N)$ } 

Consider $S^2 ( V_N)$. We list a complete set of eigenvectors,  corresponding eigenvalues with multiplicities, and the factor $\det ( 1 - x D^{ S^2(V_N) } ( g ) $ : 

{\bf Case 1:  } Different powers of the same root of unity $\omega_{  a_i } $  on the two tensor factors : 
\bea 
&& { \rm Eigenvectors } ~~~  ( | \omega_{ a_i}^{ t_1 } ; \alpha_1 \rangle \otimes | \omega_{ a_i}^{ t_2 } ; \alpha_2  \rangle   + 
| \omega_{ a_i}^{ t_2 } ; \alpha_2 \rangle \otimes | \omega_{ a_i}^{ t_1 } ; \alpha_1  \rangle  ) 
\cr 
&& ~~~~~~~~~  \alpha_1 , \alpha_2 \in \{ 1, \cdots , p_i \} ,  ~~ t_1 <  t_2 \in \{ 0, \cdots , a_i -1  \} \cr 
&& { \rm Eigenvalues } ~~~~  \omega_{ a_i}^{ t_1  + t_2 }  ~~~ { \rm Multiplicity } =  p_i^2 
\eea 
For each $a_i$ this contributes to the determinant $\det ( 1 - x D^{ S^2 (V_N) } ( \sigma ) )  $ a factor 
\bea 
F_1 (a_i )  ~~~ = ~~~ \prod_{ t_1 <  t_2  \in \{ 0, 1, \cdots , a_i -1 \}} ( 1 - x \omega_{ a_i}^{ t_1 + t_2 } )^{ p_i^2}  
\eea

{\bf Case 2: } Same power of the same root of unity on the two tensor factors with different $a_i$-subsets chosen from the $p_i$ possibilities : 
\bea 
&& { \rm Eigenvectors } ~~~~~~ (  | \omega_{ a_i}^{ t  } ; \alpha_1 \rangle \otimes | \omega_{ a_i}^{ t } ; \alpha_2  \rangle   + 
| \omega_{ a_i}^{ t } ; \alpha_2 \rangle \otimes | \omega_{ a_i}^{ t } ; \alpha_1  \rangle  ) ~~~ \cr 
&& \alpha_1 < \alpha_2 \in \{ 1 , \cdots , p_i \}  \cr 
&& {\rm Eigenvalues } ~~~~  \omega_{ a_i}^{ 2 t }  ~~~ { \rm Multiplicity } =  p_i ( p_i -1 ) /2 
\eea
This contributes to the MW-determinant a factor 
\bea 
F_2 ( a_i) = \prod_{ t = 0  }^{ a_i -1 } ( 1 - x \omega_{a_i }^{ 2 t } )^{ p_i ( p_i -1)/2} 
\eea

{\bf Case 3:} Same power of the same root of unity on the two tensor factors with same  $a_i$-subsets: 
\bea 
&& { \rm Eigenvectors } ~~~ | \omega_{ a_i}^{ t } ; \alpha  \rangle \otimes | \omega_{ a_i}^{ t } ; \alpha   \rangle 
~~~ ; ~~~~  t \in \{ 0, \cdots , a_i-1 \}   ~~~ ; ~~~ \alpha \in \{ 1, \cdots , p_i \}  \cr  
&& { \rm Eigenvalues } ~~~~  \omega_{ a_i }^{ 2 t }  ~~~ { \rm Multiplicity } =  p_i   
\eea
This contributes  a factor
 \bea 
 F_3 ( a_i) = \prod_{ t =0  }^{ a_i -1 } ( 1 - x \omega_{a_i}^{ 2 t } )^{ p_i }
 \eea

{\bf Case 4: } Finally we have the case where the two tensor factors contain eigenvectors involving different roots of unity : 
\bea 
&& { \rm Eigenvectors } ~~~ ( | \omega_{ a_i}^{ t_1 } ; \alpha_1 \rangle \otimes | \omega_{ a_j}^{ t_2 } ; \alpha_2  \rangle   + 
| \omega_{ a_j}^{ t_2 } ; \alpha_2 \rangle \otimes | \omega_{ a_i}^{ t_1 } ; \alpha_1  \rangle  ) ~~ i \ne j \in \{ 1, \cdots , K \} \cr 
&& 
~~~ t_1 \in \{ 0, 1, \cdots , a_i -1 \} , t_2 \in \{ 0 , 1, \cdots , a_j -1 \}    \cr
&&  \alpha_1 \in \{  1, \cdots , p_i \} , \alpha_2 \in \{ 1, \cdots , p_j \} \cr  
&& { \rm Eigenvalues  } ~~~  \omega_{ a_i}^{ t_1 } \omega_{a_j}^{ t_2} ~~{\rm Multiplicity }  ~~~ p_i p_j  
\eea 
This leads to the factor 
\bea 
 F_4 ( a_i , a_j )  ~~~ = ~~~ \prod_{ t_1 =1 }^{ a_i -1} \prod_{  t_2 =1 }^{ a_j-1}  ( 1 - x \omega_{ a_i}^{ t_1 } \omega_{a_j}^{ t_2 } )^{ p_1 p_2 } 
\eea

We show in Appendix  \ref{app:F-factors}
 that 
\bea\label{Feq1}  
F_1 ( a_i )  ~~~ = ~~~ \prod_{ t_1 <  t_2  \in \{ 0 , \cdots , a_i-1 \} } ( 1 - x \omega_{ a_i}^{ t_1 + t_2 } )^{ p_i^2 }  ~~~=~~~ {( 1 - x^{a_i} )^{ a_i p_i^2 \over 2  }  \over ( 1-  x^{ a_i \over G( 2,a_i )} )^{ G ( 2 , a_i ) ~ p_i^2 \over 2 }  } 
\eea
\bea \label{Feq2} 
F_2( a_i) = ~~~~ \prod_{ t=1 }^{ a_i -1 } ( 1 - x \omega_{a_i}^{ 2 t } )^{ p_i ( p_i -1)\over 2 }   
= ( 1-  x^{ a_i \over G( 2,a_i )} )^{ G ( 2 , a_i ) ~ p_i ( p_i  -1 ) \over 2 }
\eea
\bea\label{Feq3} 
F_3 ( a_i ) =   \prod_{ t =1 }^{ a_i -1 } ( 1 - x \omega_{a_i}^{ 2 t } )^{ p_i } 
   = ( 1-  x^{ a_i \over G( 2,a_i )} )^{ G( 2 , a_i ) p_i  } 
\eea
\bea\label{Feq4}  
&& F_4 ( a_i , a_j ) = \prod_{ t_1 =1}^{ a_i-1} \prod_{  t_2  =1  }^{ a_j -1}  ( 1 - x \omega_{ a_i }^{ t_1 } \omega_{a_j }^{ t_2 } )^{ p_1 p_2 }   =  \prod_{ t_1 =0  }^{a_i-1}    (  1 - x^{ a_j} \omega_{ a_i }^{ a_j t_1 } )^{ p_i p_j }  =  ( 1 - x^{ L ( a_i , a_j  )} )^{ G (  a_i ,  a_j) p_i p_j    }\cr 
&& 
\eea

Collecting terms 
\bea 
{ 1 \over \det ( 1 - x D^{ S^2(V_N) } ( \sigma ) ) }  = \prod_{ i } ( F_1 ( a_i ) F_2 ( a_i ) F_3 ( a_i )  )^{ -1} \prod_{ i < j }  ( F_4 ( a_i , a_j ) )^{ -1} 
\eea
Using the equations \eqref{Feq1}\eqref{Feq2}\eqref{Feq3} \eqref{Feq4}, we have  
\begin{equation}\label{MWS2VN}  
\boxed{ ~~~
 { 1 \over \det ( 1 - D^{ S^2(V_N) } ( \sigma ) ) } 
= \prod_{ i } \frac{ 1} { ( 1-  x^{ a_i\over G( 2,a_i)} )^{ G( 2 , a_i) {  p_i  \over 2 }} ( 1 - x^{a_i} )^{ a_i p_i^2 \over 2  }}  
\prod_{ i < j } \frac{1}{( 1 - x^{ L ( a_i , a_j )} )^{ G ( a_i ,  a_j ) p_i p_j  } }~~~
} 
\end{equation}

\subsection{ Molien-Weyl determinant for $\Lambda^2(V_N)$ } 

 We  now consider the antisymmetric squared tensor power $ \Lambda^2(V_N)$ and list a complete set of eigenvectors,  corresponding eigenvalues with multiplicities, and the factor $\det ( 1 - x D^{ \Lambda^2(V_N) } ( g ) $. 
 
 {\bf Case 1:} Different powers of the same root of unity $\omega_{  a_i } $  in the two tensor factors: 
 \bea 
&& { \rm Eigenvectors } ~~~  ( | \omega_{ a_i}^{ t_1 } ; \alpha_1 \rangle \otimes | \omega_{ a_i}^{ t_2 } ; \alpha_2  \rangle   - 
| \omega_{ a_i}^{ t_2 } ; \alpha_2 \rangle \otimes | \omega_{ a_i}^{ t_1 } ; \alpha_1  \rangle  ) 
\cr 
&& ~~~~~~~~~  \alpha_1 , \alpha_2 \in \{ 1, \cdots , p_i \} ,  ~~ t_1 < t_2 \in \{ 0, \cdots , a_i -1  \} \cr 
&& { \rm Eigenvalues } ~~~~  \omega_{ a_i}^{t_1  + t_2 }  ~~~ { \rm Multiplicity } =  p_i^2 
\eea 
For each $a_i$ this contributes to the determinant $\det ( 1 - x D^{ S^2 (V_N) } ( \sigma ) )  $ a factor 
\bea 
F_1 (a_i )  ~~~ = ~~~ \prod_{ t_1 <  t_2 \in \{ 0, 1, \cdots , a_i -1 \} } ( 1 - x \omega_{ a_i}^{ t_1 + t_2 } )^{ p_i^2}  
\eea

{\bf Case 2: } Same power of the same root of unity on the two tensor factors with different $a_i$-subsets chosen from the $p_i$ possibilities : 
\bea 
&& { \rm Eigenvectors } ~~~~~~ (  | \omega_{ a_i}^{ t  } ; \alpha_1 \rangle \otimes | \omega_{ a_i}^{ t } ; \alpha_2  \rangle   - 
| \omega_{ a_i}^{ t } ; \alpha_2 \rangle \otimes | \omega_{ a_i}^{ t } ; \alpha_1  \rangle  ) ~~~ \cr
&& \alpha_1 < \alpha_2 \in \{ 1, 2, \cdots , p_i \} \cr 
&& {\rm Eigenvalues } ~~~~  \omega_{ a_i}^{ 2 t }  ~~~ { \rm Multiplicity } =  p_i ( p_i -1 ) /2 
\eea
This contributes to the MW-determinant a factor 
\bea 
F_2 ( a_i) = \prod_{ t =1 }^{ a_i -1 } ( 1 - x \omega_{a_i }^{ 2 t } )^{ p_i ( p_i -1)/2} 
\eea

There is no ${\bf Case ~ 3 }$ or corresponding  $F_3$ factor, unlike the $S^2(V_N)$ case.

{\bf Case 4: }  Finally we have the case where the two tensor factors contain eigenvectors involving different roots of unity : 
\bea 
&& { \rm Eigenvectors } ~~~ ( | \omega_{ a_i}^{ t_1 } ; \alpha_1 \rangle \otimes | \omega_{ a_j}^{ t_2 } ; \alpha_2  \rangle   - 
| \omega_{ a_j}^{ t_2 } ; \alpha_2 \rangle \otimes | \omega_{ a_i}^{ t_1 } ; \alpha_1  \rangle  ) ~~ i \ne j \in \{ 1, \cdots , K \} \cr 
&& 
~~~ t_1 \in \{ 0, 1, \cdots , a_i -1 \} , t_2 \in \{ 0 , 1, \cdots , a_j -1 \}    \cr
&&  \alpha_1 \in \{  1, \cdots , p_i \} , \alpha_2 \in \{ 1, \cdots , p_j \} \cr  
&& { \rm Eigenvalues  } ~~~  \omega_{ a_i}^{ t_1 } \omega_{a_j}^{ t_2} ~~{\rm Multiplicity }  ~~~ p_i p_j  
\eea 
This leads to the factor 
\bea 
 F_4 ( a_i , a_j )  ~~~ = ~~~ \prod_{ t_1 =1 }^{ a_i -1} \prod_{  t_2 =1 }^{ a_j-1}  ( 1 - x \omega_{ a_i}^{ t_1 } \omega_{a_j}^{ t_2 } )^{ p_1 p_2 } 
\eea 

Collecting the different cases, for a permutation $\sigma $ with cycle structure $[a_1^{ p_1} , a_2^{ p_2} , \cdots , a_K^{ p_K} ]$: 
\bea
\det ( 1 - D^{ \Lambda^2 ( V_N ) } ( \sigma )  = \prod_{ i } F_1 ( a_i ) F_2 ( a_i ) \prod_{ i <  j } F_4 ( a_i , a_j ) 
\eea
Using the equations \eqref{Feq1},\eqref{Feq2}, \eqref{Feq4},  the Molien-Weyl generating function for invariants is : 
\begin{equation}\label{MWL2N} 
\boxed{ 
 { 1 \over \det ( 1 -  x D^{ \Lambda^2 ( V_N ) } ( \sigma ) ) }  
 =  \prod_{ i } \frac{( 1 - x^{ a_i \over G( 2, a_i ) })^{{ p_i \over 2 }  G( 2 , a_i ) }}{( 1 - x^{ a_i} )^{ a_i p_i^2 \over 2 } } \prod_{ i < j } \frac{1}{( 1 - x^{ L ( a_i , a_j )} )^{ G ( a_i,  a_j)  p_i p_j   } }
 } 
\end{equation} 
The equations \eqref{MWL2N}, \eqref{MWS2VN} and \eqref{MainProp} show that 
\bea 
 { 1 \over \det ( 1 - x D^{ \Lambda^2 ( V_N ) } ( \sigma ) ) }  { 1 \over \det ( 1 -x  D^{ S^2(V_N) } ( \sigma ) ) }  
 =  { 1 \over \det ( 1 - x D^{ ( V_N \otimes  V_N ) } ( \sigma ) ) } 
\eea
which is as expected since 
\bea 
 \det ( 1 - x D^{ ( V \oplus W  ) } ( \sigma )  )  =  \det ( 1 - x D^{ ( V  ) } ( \sigma )  )  \det ( 1 - x D^{ (  W  ) } ( \sigma )  ) 
\eea
and $ V_N \otimes V_N  = S^2 ( V_N ) \oplus \Lambda^2 ( V_N )$.

\subsection{ Molien-Weyl determinants for the irreps $V_2 , V_3 $ } 

Using the decomposition into irreducibles obtained as equation \eqref{decompL2N} from section \ref{sec:symmirreps}
\bea 
\Lambda^2 ( V_N ) = V_H \oplus V_{ 3 }  
\eea
we have 
\bea 
\det ( 1  - x D^{ \Lambda^2 ( V_N )  } ( \sigma )  ) =  \det ( 1  - x D^{ V_H } ( \sigma ) )  \det (  1 - x D^{ V_3 } ( \sigma )  )
\eea
Hence 
\bea 
{ 1 \over \det (  1 - x D^{ V_3 } ( \sigma )  ) } =
{  \det ( 1  - x D^{ V_H } ( \sigma )  )  \over \det ( 1  - x D^{ \Lambda^2 ( V_N )  } ( \sigma )  ) } 
\eea 
Therefore, for $\sigma $ having a general  cycle structure $[ a_1^{ p_1} , a_2^{ p_2} , \cdots , a_K^{ p_K} ] $, using 
\eqref{MWVH} and \eqref{MWL2N}, we have: 
\bea 
&& { 1 \over \det (  1 - x D^{ V_3 } ( \sigma ) )  } = 
{ 1 \over ( 1 - x ) } \prod_{ i } ( 1 - x^{ a_i } )^{ p_i} \times  \prod_{ i } \frac{( 1 - x^{ a_i \over G( 2, a_i ) })^{{ p_i \over 2 }  G( 2 , a_i ) }}{( 1 - x^{ a_i} )^{ a_i p_i^2\over  2  }}
\cr 
&& \hspace*{5cm} ~~\times ~~ \prod_{ i < j } \frac{1}{( 1 - x^{ L ( a_i , a_j )} )^{ a_i a_j p_i p_j \over L ( a_i , a_j )  } }
\eea

To obtain the Molien-Weyl determinant for $V_2$ we use the decomposition \eqref{decompS2N} from section \ref{sec:symmirreps}
\bea 
S^2 ( V_N ) = 2V_0 \oplus 2V_H \oplus V_2 
\eea
We have 
\bea
&& \det ( 1 - x D^{ S^2( V_N ) } ( \sigma ) ) = ( \det ( 1 - x D^{ V_0 } ( \sigma ) ))^2 
( \det ( 1 - x D^{ V_H } ( \sigma ) ))^2 ( \det ( 1 - x D^{ V_2  } ( \sigma ) )) \cr 
&& 
\eea
This allows us to write 
$ ( \det ( 1 - x D^{ V_2  } ( \sigma ) ))^{ -1}  $ in terms of determinants, notably \eqref{MWVN} and \eqref{MWS2VN}, which  we have already calculated : 
\bea 
&& ( \det ( 1 - x D^{ V_2  } ( \sigma ) ))^{-1}  \cr 
&&  =( \det ( 1 - x D^{ V_0 } ( \sigma ) ))^2 
( \det ( 1 - x D^{ V_H } ( \sigma ) ))^2  \det ( 1 - x D^{ S^2( V_N ) } ( \sigma ) )^{-1}  \cr 
&& = ( \det ( 1 - x D^{ V_N } ( \sigma ) ))^2  \det ( 1 - x D^{ S^2( V_N ) } ( \sigma ) )^{-1}  \cr 
&&    =  \left (  \prod_i (  1 - x^{ a_i}  )^{ 2 p_i } \right ) \det ( 1 - x D^{ S^2( V_N ) } ( \sigma ) )^{-1}   \cr 
&& = \prod_i (  1 - x^{ a_i}  )^{ 2 p_i }\prod_{ i } \frac{1}{  ( 1-  x^{ a_i\over G( 2,a_i)} )^{ {  p_i  \over 2 }G ( 2 , a_i)} { ( 1 - x^{a_i} )^{ a_i p_i^2 \over 2 }} }  
\prod_{ i < j } \frac{1}{( 1 - x^{ L ( a_i , a_j )} )^{ G ( a_i , a_j)  p_i p_j   } } \cr 
&& 
\eea

\subsection{ Physical partition function }

We can now use the results for the Molien-Weyl determinants obtained above to make the formula \eqref{proddetsref}   explicit.
It is convenient  to define $ x_1 = e^{ - \beta_0^1} , x_2 = e^{ - \beta^2_0}$ for the irrep $V_0$, $ x_3 = e^{ - \beta_H^{ 1} } , x_4 =e^{ - \beta_H^{ 2}}, x_5 =e^{ - \beta_H^{ 3 } } $ for the irrep $V_H$, $x_6 = e^{ - \beta_2 } $ For $V_2$ and $ x_7 = e^{ - \beta_3}$ for $V_3$. 
The product of Molien-Weyl determinants with the respective $x_i$' s is 
\bea 
&&  { 1 \over \det ( 1 - x_1 D^{ V_0}  ( \sigma ) ) } { 1 \over \det ( 1 - x_2 D^{ V_0} ( \sigma ) )  } 
{ 1 \over \det ( 1 - x_3 D^{ V_H} ( \sigma ) ) } { 1 \over \det ( 1 - x_4 D^{ V_H } ( \sigma ) )  } { 1 \over \det ( 1 - x_5 D^{ V_H}  ( \sigma ) ) } \cr 
&&\times  { 1 \over \det ( 1 - x_6 D^{ V_2}   ( \sigma ) )  } { 1 \over \det ( 1 - x_7 D^{ V_3 } ( \sigma )  ) }  
\eea
For $ [ \sigma ] = [ a_1^{ p_1} , a_2^{ p_2} , \cdots , a_K^{ p_K} ]   $
\bea\label{refinedpdep}  
&&    \cZ (N  ,  p ; x_1 , \cdots , x_7  ) \cr 
  && = { 1 \over \det ( 1 - x_1 D^{ V_0} ( \sigma )) } { 1 \over \det ( 1 - x_2 D^{ V_0} ( \sigma ) ) } 
{ 1 \over \det ( 1 - x_3 D^{ V_H} ( \sigma )  ) } { 1 \over \det ( 1 - x_4 D^{ V_H } ( \sigma ) )} { 1 \over \det ( 1 - x_5 D^{ V_H}  ( \sigma ) ) }   \cr
&& { 1 \over \det ( 1 - x_6 D^{ V_2}   ( \sigma )) } { 1 \over \det ( 1 - x_7 D^{ V_3 } ( \sigma )  ) }   \cr 
&& = {  (  1 - x_3 ) ( 1- x_4 ) ( 1- x_5 ) \over ( 1 - x_1 ) ( 1 - x_2 )}  \prod_{ i } { 1 \over ( 1 - x_3^{ a_i} )^{ p_i } } { 1 \over ( 1 - x_4^{ a_i} )^{ p_i } } { 1 \over ( 1 - x_5^{ a_i} )^{ p_i } } \cr  
&& ~~~ \times ~~~ \prod_i (  1 - x_6^{ a_i}  )^{ 2 p_i }\prod_{ i } \frac{1}{ ( 1-  x_6^{ a_i\over G( 2,a_i)} )^{  {  p_i  \over 2 } G ( 2 , a_i)} { ( 1 - x_6^{a_i} )^{ a_i p_i^2 \over 2  }}   }
\prod_{ i < j } \frac{1}{( 1 - x_6^{ L ( a_i , a_j )} )^{ G ( a_i,  a_j )  p_i p_j   } } \cr 
&& ~~~ \times ~~~~{ 1 \over ( 1 - x_7 ) } \prod_{ i } ( 1 - x_7^{ a_i } )^{ p_i}  \prod_{ i } \frac{( 1 - x_7^{ a_i \over G( 2, a_i ) })^{{ p_i \over 2 }  G ( 2 , a_i ) }}{( 1 - x_7^{ a_i} )^{ a_i p_i^2 \over 2 } } \prod_{ i < j } \frac{1}{( 1 - x_7^{ L( a_i , a_j )} )^{ G ( a_i ,  a_j)  p_i p_j }} \cr 
&& 
\eea
The general invariant partition function is 
\bea\label{generalAns}  
 \cZ (N ;  x_1, \cdots , x_7 )  = \sum_{ p \vdash N  } { 1 \over \Sym ~ p } \cZ ( N , p ; x_1 , \cdots , x_7  )  
\eea
Note that when we set $ x_i \rightarrow x $, the expression simplifies to \eqref{MainProp} and we recover, as expected,  the partition function of the unrefined theory .

\section{ Summary and  Outlook  }

In this paper we have calculated the partition function for matrix quantum mechanics with gauged permutation symmetry. We found explicit formulae for the case of harmonic oscillator systems as a sum over partitions of $N$, where the summands depend on elementary number theoretic functions of the partitions. The phase structure of these partition functions will be studied in detail in   upcoming work 
\cite{GPIMQM-Thermo}. The rapid growth of the number of invariant states as a function of the number of oscillators (as evident from the sequences given in \cite{LMT} and \cite{oeis1})  leads us to expect that the partition function given in \eqref{partitionSum} and \eqref{MainProp} will have the characteristics of a system with vanishing Hagedorn temperature at large $N$ similar to what is encountered in tensor quantum mechanics with continuous symmetries  \cite{Tseyt,Kleb} related to a similar growth of the number of states \cite{JBGSR1,JBGSR2,oeis1,oeis2}. This is to be contrasted with the finite Hagedorn temperature observed in   multi-matrix systems
\cite{Aha2Mat}, which has been revisited with an interpretation in terms of small black holes in AdS \cite{HanMaltz,BerSmall,BerPart}.  The finite $N$ formulae for both canonical and micro-canonical ensembles available in the systems studied here make these a very interesting set-up to study  the approach to the transition at large $N$ and the relation between the behaviours of the canonical and micro-canonical ensembles in this approach.  Finite group gauge symmetries have been considered in lattice gauge theory
\cite{Rebbi-1980,Bhanot-Rebbi-1981,HN2001,BC2020}   and comparisons between the phase structure of the quantum mechanical models at hand with phases encountered in lattice gauge theory form additional motivation for the thermodynamic investigations of the partition functions derived here. 

It is also worth noting that the canonical partition functions studied here (equations \eqref{partitionSum}  \eqref{LCMformula} \eqref{generalAns} \eqref{refinedpdep}) can be summed up into rational functions of the form $ P (x)/Q(x) $ and as such define Hilbert series for commutative rings. The denominators encode information about the generators of the ring. 
They have a graph-theoretic interpretation, as  observed and illustrated in \cite{LMT}, and as studied in more detail in \cite{PIG2MM}. The numerators encode relations between the generators, as well as higher order relations. Applications of Hilbert series in connection with group invariants with physical applications are an active area of research in particle physics and string theory (see e.g. 
\cite{Jenkins:2009dy}\cite{Hanany:2008sb}\cite{Henning:2017fpj}\cite{deMelloKoch:2022dpj}). 
The structure of these rational forms for the partition functions given here, and the elucidation of the generators and relations encoded, is an interesting subject of further study. 

The group-theoretic solutions to the 13-parameter Gaussian matrix model \cite{PIGMM} (and related models for symmetric matrices having vanishing diagonals \cite{PIGMFC}, multi-matrix models \cite{PIG2MM} and tensor models \cite{PIGTM})  along with  the calculations of partition functions using techniques developed here  and the path integral formulation in \cite{GPIMQM-PI} set up the framework for the study of interesting perturbations of these models by higher order permutation invariant  interactions. Any permutation invariant polynomial function of degree greater than two can be viewed as an interaction term. In the matrix data science applications
of zero-dimensional matrix models such polynomials of degree 3 and 4 have been used to detect and rank the strengths of non-Gaussianities  real-world matrix ensembles \cite{GTMDS}\cite{CHRS}\cite{PIGMFC}. Adding the higher ranked polynomial functions to matrix models and performing perturbative calculations of correlators will be useful in these applications. Analogous calculations in the matrix quantum systems will allow us to investigate the effect of interactions on thermodynamic features. It will be interesting to investigate the existence of  double scaling limits analogous to the melonic limits of tensor models \cite{Gur2010,GurRiv2011}.  The phase structures of the perturbed models should be accessible to hybrid Monte Carlo methods which have been applied to multi-matrix systems in the context of BFSS \cite{bfss} and BMN \cite{bmn}  models \cite{Berkowitz:2016jlq,Catterall:2010gf,Asano:2018nol,Pateloudis:2022oos}.

\vskip.3cm

\centerline{\bf{Acknowledgments}}
\vskip.2cm 

DO'C  is supported by the Irish Research Council and Science Foundation
Ireland under grant SFI-IRC-21/PATH-S/9391.
SR is supported by the Science and Technology Facilities Council (STFC) Consolidated
Grants ST/P000754/1 “String theory, gauge theory and  duality” and ST/T000686/1
“Amplitudes, strings and  duality”. SR thanks the Dublin Institute for Advanced Studies for hospitality while part of this work was being done. 
We thank   Yuhma Asano, George Barnes,  David Berenstein, Veselin Filev, Robert de Mello Koch,  Antal  Jevicki,  Adrian Padellaro for  discussions related to the subject of this paper.

\vskip.5cm 

\begin{appendix}

\section{ Computation of products over roots of unity  } 
\label{app:F-factors}

Here we calculate 
\bea 
F_1 ( a_i )  ~~~ = ~~~ \prod_{ t_1  < t_2  \in \{ 0 , \cdots , a_i-1 \} } ( 1 - x \omega_{ a_i}^{ t_1 + t_2 } )^{ p_i^2} 
\eea 
We first calculate 
\bea 
  \hat F_1 ( a_i )  \equiv \prod_{ t_1 <  t_2  \in \{ 0 , \cdots , a_i-1 \} } ( 1 - x \omega_{ a_i}^{ t_1 + t_2 } )  
\eea
Note that  $ ( 1 - x \omega_{ a_i}^{ t_1 + t_2 } )   $ is unchanged when we swop $t_1 , t_2 $. This means 
\bea 
  ( \hat F_1 ( a_i )  )^2 = \prod_{ t_1 \ne   t_2  \in \{ 0 , \cdots , a_i-1 \} } ( 1 - x \omega_{ a_i}^{ t_1 + t_2 } )  
\eea
This can be calculated as follows : 
\bea 
  ( \hat F_1 ( a_i )  )^2
& = &  \prod_{ t_1 =1 }^{ a_i-1} \prod_{ t_2 \ne t_1 } ( 1 - x \omega_{ a_i}^{ t_1 + t_2 } )  \cr 
&= & \prod_{ t_1 =1 }^{ a_i-1}  { 1 \over ( 1 - x \omega_{ a_i}^{ 2t_1 }   ) } \prod_{ t_2 =0 }^{ a_i -1}  ( 1 - x \omega_{ a_i}^{ t_1 + t_2 } )  \cr 
& = &  \prod_{ t_1}  { ( 1 - x^{ a_i }  \omega_{ a_i}^{ t_1 a_i} ) \over ( 1 - x \omega_{ a_i}^{ 2t_1 }   ) }
\eea
where we used Lemma 1 (equation  \eqref{prodroots}) for the product over $t_2$. Simplifying further 
\bea 
&& ( \hat F_1 ( a_i ) )^2   =  \prod_{ t_1 =0 }^{ a_i -1 }   { ( 1 - x^{ a_i }  \omega_{ a_i}^{ t_1 a_i} ) \over ( 1 - x \omega_{ a_i}^{ 2t_1 }   ) } 
= \prod_{ t_1}  { ( 1 - x^{ a_i }   ) \over ( 1 - x \omega_{ a_i}^{ 2t_1 }   ) }  = ( 1 - x^{ a_i }   )^{ a_i }  \prod_{ t_1}  { 1 \over ( 1 - x \omega_{ a_i}^{ 2t_1 }   ) } 
\eea 
 If $a_i$ is odd, $ \omega_{ a_i}^2 $ is a primitive $a_i$'th root of unity and we have (using Lemma 1 ) : 
\bea 
\prod_{ t_1=0}^{ a_i -1 }    {  1 \over ( 1 - x \omega_{ a_i}^{ 2  t_1   }   ) } = { 1 \over ( 1 - x^{ a_i} ) }
\eea
If $a_i  $ is even we can write $t_1  = { a_i \over 2}  r  + s  $ with $ r \in \{ 0 , 1 \} , s \in \{ 0 , \cdots , {a_i \over 2 } -1 \}$ : 
\bea 
&& \prod_{ t_1 =1 }^{ a_i -1} {  1 \over ( 1 - x \omega_{ a_i}^{ 2  t_1   }   ) }
= 
\prod_{ r \in \{ 0 , 1  \} } \prod_{ s = 0 }^{ {a_i \over 2}  -1 }  { 1 \over ( 1 - x \omega_{ a_i}^{ a_i r + 2 s } ) } \cr 
&& 
= \prod_{ r \in \{ 0 , 1  \} } \prod_{ s = 0 }^{ {a_i \over 2}  -1 } { 1 \over ( 1 - x \omega_{ { a_i \over 2}  }^{  s} ) }\cr 
&& 
= \prod_{ r \in \{ 0 , 1 \} } { 1 \over ( 1 - x^{ a_i \over 2} ) } = { 1 \over ( 1 - x^{ a_i  \over 2 } )^2 } 
\eea
The odd and even cases  can be collected by expressing the result in terms of $\hbox{GCD}( 2 , a_i )\equiv G ( 2 , a_i ) $ which is $2$ if $a_i $ is even, and $1$ if $a_i $ is odd. Hence 
\bea\label{handyid}  
\prod_{ t_1 = 0 }^{ a_i -1 }  {  1 \over ( 1 - x \omega_{ a_i }^{ 2  t_1   }   ) } = { 1 \over ( 1-  x^{ a_i \over G( 2,a_i )} )^{ G(2 , a_i ) }  } 
\eea
We conclude that 
\bea 
 \prod_{ t_1 \ne t_2 \in \{ 0 , \cdots , a_i -1 \}   } ( 1 - x \omega_{ a_i}^{ t_1 + t_2 } ) 
~=~  { ( 1 - x^{a_i} )^{ a_i}  \over ( 1-  x^{ a_i\over G( 2,a_i)} )^{ G ( 2 , a_i) }  } 
\eea
and 
\bea 
 \hat F_1 ( a_i )  = { ( 1 - x^{a_i} )^{ a_i \over 2 }  \over ( 1-  x^{ a_i\over G( 2,a_i)} )^{ G ( 2 , a_i) \over 2  }  } 
\eea

\bea 
F_1 ( a_i) ~  = ~ 
\prod_{ t_1 <  t_2 \in \{ 0, \cdots , a_i -1 \} } ( 1 - x \omega_{ a_i}^{ t_1 + t_2 } )^{ p_i^2}  
= { ( 1 - x^{a_i} )^{ a_i p_i^2 \over 2  }  \over ( 1-  x^{ a_i \over G( 2,a_i )} )^{  p_i^2 G ( 2 , a_i )  \over 2  }  } 
\eea

The factors $F_2 ( a_i) , F_3 ( a_i) $ can be computed using \eqref{handyid}. They are 
\bea 
F_2( a_i ) = ~~~~ \prod_{ t =1 }^{ a_i  -1 } ( 1 - x \omega_{a_i }^{ 2 t  } )^{ p_i  ( p_i  -1)/2}   
~ = ~  ( 1-  x^{ a_i \over G( 2,a_i )} )^{ G( 2 , a_i ) ~ p_i  ( p_i  -1 )/2 }
\eea
and 
\bea 
F_3 ( a_i ) =   \prod_{ t =1 }^{ a_i  -1 } ( 1 - x \omega_{a_i }^{ 2 t  } )^{ p_i  } 
  ~ = ~ ( 1-  x^{ a_i \over G( 2,a_i )} )^{ G( 2 , a_i ) p_i  } 
\eea
Collecting the factors 
\bea 
F_1 ( a_i  ) F_2 ( a_i ) F_3 ( a_i )~  =~ { ( 1 - x^{a_i } )^{ a_i  p_i^2\over 2  } ( 1-  x^{ a_i \over G( 2,a_i )} )^{{  p_i   \over 2 } G ( 2 , a_i ) }    }     
\eea
Now consider the factor that depends on $a_i , a_j $, which is simplified using \eqref{prodroots} and   \eqref{RLCML} 
\bea 
&& F_4 ( a_i , a_j ) = \prod_{ t_1 ,  t_2 } ( 1 - x \omega_{ a_i }^{ t_1 } \omega_{a_j }^{ t_2 } )^{ p_i p_j  }   \cr 
&& = \prod_{ t_1 }  (  1 - x^{ a_2} \omega_{ a_1}^{ a_2 t_1 } )^{ p_i p_j } \cr 
&& =  ( 1 - x^{ L ( a_i , a_j  )} )^{ a_i  a_j  p_i p_j \over L ( a_i , a_j )   } \cr 
&& = ( 1 - x^{ L ( a_i , a_j  )} )^{ G ( a_i ,  a_j )   p_i p_j   } 
\eea

\end{appendix}

\end{document}